\documentclass[onecolumn,usenatbib]{mn2e}
\usepackage{amsmath}
\usepackage{amssymb}
\usepackage{graphicx}
\usepackage{psfig}
\usepackage{amssymb}
\usepackage{subfigure}
\usepackage{graphicx,natbib,rotating}
\def\spose#1{\hbox to 0pt{#1\hss}}
\def\approxlt{\mathrel{\spose{\lower 3pt\hbox{$\sim$}}
        \raise 2.0pt\hbox{$<$}}}
\def\approxgt{\mathrel{\spose{\lower 3pt\hbox{$\sim$}}
        \raise 2.0pt\hbox{$>$}}}
\def\approxpropto{\mathrel{\spose{\lower 3pt\hbox{$\sim$}}
        \raise 2.0pt\hbox{$\propto$}}}
\mathchardef\twiddle="2218

\def\multleft#1{\hbox to size{\vbox {\halign {\lft{##}\cr #1}}\hfill}\par}
\def\multright#1{\hbox to size{\vbox {\halign {\rt{##}\cr #1}}\hfill}\par}

\def\today{\ifcase\month\or January\or February\or March\or April\or May\or
      June\or July\or August\or September\or October\or November\or December\fi
      \space\number\day, \number\year}
\def\<{\thinspace}
\newcommand{\beq}{\begin{equation}}
\newcommand{\eeq}{\end{equation}}
\def\lsim{\mathrel{\mathpalette\@versim<}}
\def\gsim{\mathrel{\mathpalette\@versim>}}

\def\nuLupeak{\nu{L{\nu_{\rm peak}}}}
\def\nupeak{ \nu_{\rm peak}}

\def\Lubol{L_{\rm bol}}
\def\LuEdd{L_{\rm Edd}}
\def\Luline{L_{\rm BLR}}

\def\Halpha{\rm H{\alpha}}
\def\Hbeta{\rm H{\beta}}
\def\Mgii{\rm Mg~ II}
\def\Civ{\rm C~ IV}
\def\aap{\rm A~\&~A}

\def\apj{\rm ApJ}
\def\apjs{\rm ApJS}
\def\aj{\rm AJ}
\def\pasp{\rm PASP}
\def\mnras{\rm MNRAS}
\def\chjaa{\rm ChJAA}
\title[FSRQs from SDSS DR3 Quasar Catalogue]{Flat-Spectrum Radio Quasars from SDSS DR3 Quasar Catalogue}
\author[Z. Y. Chen, M. F. Gu and X. Cao]
{Zhaoyu
Chen$^{1,2}$\thanks{E-mail:zychen@shao.ac.cn},
 Minfeng Gu$^{1}$, Xinwu Cao$^{1}$\\
$^{1}$ Key Laboratory for Research in Galaxies and Cosmology,
Shanghai Astronomical Observatory, Chinese Academy of Sciences, 80\\
Nandan Road, Shanghai 200030, China\\
$^{2}$ Graduate School of the Chinese Academy of Sciences, Beijing
100039, China}
\begin{document}
\maketitle

\begin{abstract}
\noindent We constructed a sample of 185 Flat Spectrum Radio Quasars
(FSRQs) by cross-correlating the Shen et al.'s SDSS DR3 X-ray quasar
sample with FIRST and GB6 radio catalogues. From the spectrum energy
distribution (SED) constructed using multi-band (radio, UV, optical,
Infrared and X-ray) data, we derived the synchrotron peak frequency
and peak luminosity. The black hole mass $M_{\rm BH}$ and the broad
line region (BLR) luminosity (then the bolometric luminosity $L_{\rm
bol}$) were obtained by measuring the line-width and strength of
broad emission lines from SDSS spectra. We define a subsample of 118
FSRQs, of which the nonthermal jet emission is thought to be
dominated over the thermal emission from accretion disk and host
galaxy. For this subsample, we found 25 FSRQs having synchrotron
peak frequency $\nupeak
> 10^{15}$ Hz, which is higher than the typical value for FSRQs.
These sources with high $\nupeak$ could be the targets for the Fermi
Gamma-ray telescope. Only a weak anti-correlation is found between
the synchrotron peak frequency and peak luminosity, while no strong
correlation is present either between the synchrotron peak frequency
and black hole mass, or between the synchrotron peak frequency and
the Eddington ratio $\Lubol/\LuEdd$. When combining the FSRQs
subsample with the Wu et al.'s sample of 170 BL Lac objects, the
strong anti-correlation between the synchrotron peak frequency and
luminosity apparently presents covering about seven order of
magnitude in $\nupeak$. However, the anti-correlation differs with
the blazar sequence in the large scatter. At similar peak frequency,
the peak luminosity of FSRQs with $\nupeak > 10^{15}$ Hz is
systematically higher than that of BL Lac objects, with some FSRQs
out of the range covered by BL Lac objects. Although high $\nupeak$
are found in some FSRQs, they do not reach the extreme value of BL
Lacs. For the subsample of 118 FSRQs, we found significant
correlations between the peak luminosity and black hole mass, the
Eddington ratio, and the BLR luminosity, indicating that the jet
physics may be tightly related with the accretion process.

\end{abstract}


\section{Introduction}

Blazars, including BL Lac objects and flat-spectrum radio quasars
(FSRQs), are the most extreme class of active galactic nuclei
(AGNs), characterized by strong and rapid variability, high
polarization, and apparent superluminal motion. These extreme
properties are generally interpreted as a consequence of non-thermal
emission from a relativistic jet oriented close to the line of
sight. As such, they represent a fortuitous natural laboratory with
which to study the physical properties of jets, and, ultimately, the
mechanisms of energy extraction from the central supermassive black
holes.

The most prominent characteristic of the overall spectral energy
distribution (SED) of blazars is the double-peak structure with two
broad spectral components. The first, lower frequency component is
generally interpreted as being due to synchrotron emission, and the
second, higher frequency one as being due to inverse Compton
emission. BL Lac objects (BL Lacs) usually have no or only very weak
emission lines, but have a strong highly variable and polarized
non-thermal continuum emission ranging from radio to ${\rm
\gamma}$-ray band, and their jets have synchrotron peak frequencies
ranging from IR/optical to UV/soft-X-ray energies. Compared to BL
Lacs, FSRQs have strong narrow and broad emission lines, however
generally have low synchrotron peak frequency. According to the
synchrotron peak frequency, BL Lac objects can be divided into three
subclasses, i.e. low frequency peaked BL Lac objects (LBL),
intermediate objects (IBL) and high frequency peaked BL Lac objects
(HBL) \citep{padovani95}. In general, radio-selected BL Lacs tend to
be LBLs, and XBLs are HBLs \citep{urry95}. \cite{fossati98} and
\cite{ghisel98} have proposed the well-known `blazar sequence' ,
which plot various powers vs the synchrotron peak frequency
$\nupeak$ including FSRQs and BL Lacs. They draw a conclusion that
the peak frequency seem to be anti-correlated with the source power
with the most powerful sources having the relatively small $\nupeak$
and the least powerful ones having the highest $\nupeak$.
\cite{ghisel98} gave a theoretical interpretation to these
anti-correlations, namely, the more powerful sources suffered a
larger probability of losing energy, the more cooling the sources
subjected, thus translates into a lower value of $\nupeak$.

However, recently the blazar sequence has been largely in
debates. %$^{[6,7]}$. %Especially, the blue quasars were found by
%Padovani et al. $^{[?]}$, which is thought to be the outliers of
%blazar sequence.
%\cite{fosch06} performed a homogeneous and systematic analysis of
%simultaneous X-ray and optical/UV properties of a group of 15 ${\rm
%\gamma}$-ray loud AGN, which include 13 blazars (6 BL Lacs and 7
%FSRQs) and 2 radio galaxies. They found all the selected sources
%appear to follow the classic `blazar sequence'.
By constructing a large sample of about 500 blazars from the Deep
X-Ray Radio Blazar Survey (DXRBS) and the ROSAT All-Sky Survey-Green
Bank Survey (RGB), \cite{padovani03} found that the ``X-ray-strong''
radio quasars, with similar SED to that of HBLs, have much higher
synchrotron peak frequencies than those of classical FSRQs. Their
DXRBS sample does not show the expected blazar sequence. Exceptions
to blazar sequences were also found by \cite{anton05} that most of
the low radio luminosity sources have synchrotron peaks at low
frequencies, instead of the expected high frequencies in their
blazar sample. They claimed that at least part of the systematic
trend seen by \cite{fossati98} and \cite{ghisel98} results from
selection effects. \cite{nieppola06} have studied a sample of over
300 BL Lacs objects, of which 22 objects have high $\nupeak>10^{19}$
Hz. There are negative correlations between $\rm \nupeak$ and the
luminosity at 5 GHz, 37 GHz, and 5500$\rm \AA$, however, the
correlation turns to slightly positive in X-ray band. Moreover, they
claimed that there is no significant correlation between source
luminosity at synchrotron peak frequency and $\nupeak$, and several
low energy peaked BL Lacs with low radio luminosity were also found.
By using the results of very recent surveys, \cite{padovani07}
claimed that there is no anti-correlation between the radio power
and synchrotron peak frequency in blazars once selection effects are
properly taken into account, and some blazars were found to have low
power as well as low $\nupeak$, or high power and high $\nupeak$ as
well. Furthermore, FSRQs with synchrotron peak frequency in the
UV/X-ray band have been claimed \cite[e.g.][]{gio07,pad02}. In
summary, it seems that the blazar sequence in its simplest form
cannot be valid \citep{padovani07}.
%Ten FSRQs as candidate sources to have an X-ray spectrum dominated
%by jet synchrotron emission, however are proven to have thermal
%accretion disk emission in X-ray band \citep{landt08}.

In this paper, we investigate the dependence of synchrotron peak
frequency on the peak luminosity for a sample of FSRQs selected from
SDSS DR3 quasar catalogue. The advantage of our sample is that the
SDSS spectra enable us to measure the various broad emission lines,
and then to estimate the black hole mass and the broad line region
luminosity. Therefore, the relationship between the synchrotron peak
frequency and black hole mass and Eddington ratio can be explored,
which might have potential importance for us to study the jet
formation and physics and the jet-disk relation. We present the
sample selection in $\S$ 2. The reduction of SDSS spectra and the
estimation of black hole mass and broad line region luminosity are
described in $\S$ 3. The derivation of synchrotron peak frequency
and luminosity are given in $\S$ 4, in which the thermal emission
from accretion disc and host galaxy are also calculated. The various
correlation analysis are shown in $\S$ 5, of which we focus on the
relationship between the synchrotron peak frequency and luminosity.
$\S$ 6 is dedicated to discussions. Finally, the summary is given in
$\S$ 7. The cosmological parameters ${\rm H_0=70 \,
km~s^{-1}~Mpc^{-1}}$, $\Omega_{\rm m}$=0.3, $\Omega_{\Lambda}$ = 0.7
are used throughout the paper, and the spectral index $\alpha$ is
defined as $f_{\nu}$ $\propto$ $\nu^{-\alpha}$ with $f_{\nu}$ being
the flux density at frequency $\nu$.

\section{The sample selection}

\subsection{The SDSS DR3 X-ray quasar sample}

We started from the SDSS DR3 X-ray quasar sample of \cite{Shen06},
which is the result of individual X-ray detections of SDSS DR3
quasar catalogue (Schneider et al. 2005) in the images of ROSAT All
Sky Survey (RASS). The SDSS DR3 quasar catalogue consists of 46,420
objects with luminosities brighter than $M_{i}=-22$, with at least
one emission line with full width at half-maximum (FWHM) larger than
1000 $\rm km~ s^{-1}$ and with highly reliable redshifts. A few
unambiguous broad absorption line quasars are also included. The sky
coverage of the sample is about 4188 $\rm deg^2$ and the redshifts
range from 0.08 to 5.41. The five-band $(u,~ g,~ r,~ i,~ z)$
magnitudes have typical errors of about 0.03 mag. The spectra cover
the wavelength range from 3800 to 9200 $\rm \AA$ with a resolution
of about 1800 - 2000 (see Schneider et al. 2005 for details). The
soft X-ray properties of SDSS DR3 quasars have been investigated by
\cite{Shen06} through individual detection and stacking analysis.
\cite{Shen06} applied the upper-limit maximum likelihood method to
detect the X-ray flux at the position of each SDSS DR3 quasar and
accept the objects with detection liklihood $L>7$ as individual
detections. The number of these individual X-ray detected quasars
were 3366, which is about 25 percent higher than the RASS catalogue
matches \cite[see][for details]{Shen06}. The 1 keV X-ray luminosity
in the source rest frame of this 3366 X-ray quasar sample is
obtained by assuming a power-law distribution of X-ray photons ${\rm
N(E) \propto E^{-\Gamma}}$ with $\Gamma\sim2$ and corrected for
absorption using the fixed column density at the Galactic value
according to \cite{dickey90} for each source \citep{Shen06}.

\subsection{Cross-correlation with FIRST and GB6 radio catalogues}

%adding description of first, sdss, gb6 and 2mass introduction,
%positional uncertainty and the matching radius.
In this paper, we define a quasar to be FSRQ according to the radio
spectral index. Therefore, we cross-correlate the SDSS DR3 X-ray
quasar sample with Faint Images of the Radio Sky at
Twenty-Centimeters 1.4 GHz radio catalogue \cite[FIRST, ][]{bec95}
and the Green Bank 6 cm radio survey at 4.85 GHz radio catalogue
\cite[GB6, ][]{gre96}, which are two of the largest radio surveys
well matched with SDSS sky coverage. The FIRST survey used the VLA
to observe the sky at 20 cm (1.4 GHz) with a beam size of
$5.^{''}4$.
%and an rms sensitivity of about 0.15 mJy $\rm beam^{-1}$.
FIRST was designed to cover the same region of the sky as the SDSS,
and it observed 9000 $\rm deg^{2}$ at the north Galactic cap and a
smaller $\sim2.5^{\circ}$ wide strip along the celestial equator. It
is 95\% complete to 2 mJy and 80\% complete to the survey limit of 1
mJy. The survey contains over 800,000 unique sources, with an
astrometric uncertainty of $\lesssim 1^{''}$.
%FIRST
%includes two measures of 20 cm continuum flux density: the peak
%value, Fpeak, and the integrated flux density, Fint, measured by
%fitting a two-dimensional Gaussian to coadded images of each source.
%The FIRST survey area has been chosen to coincide with that of the
%Sloan Digital Sky Survey (SDSS).
Due to the deeper survey limit and higher resolution, we prefer to
use FIRST, instead of NVSS, which was also carried out using the VLA
at 1.4 GHz to survey the entire sky north of $\delta=-40^{\circ}$
and contains over 1.8 million unique detections brighter than 2.5
mJy, however with lower spatial resolution $\rm 45^{''} beam^{-1}$.
The GB6 survey at 4.85 GHz was executed with the 91 m Green Bank
telescope in 1986 November and 1987 October. Data from both epochs
were assembled into a survey covering the $0^{\circ} < \delta <
75^{\circ}$ sky down to a limiting flux of 18 mJy, with $3.5^{'}$
resolution. GB6 contains over 75,000 sources, and has a positional
uncertainty of about $10^{''}$ at the bright end and about $50^{''}$
for faint sources \citep{kim08}.

The sample of 3366 quasars was firstly cross-correlated between the
SDSS quasar positions and the FIRST catalogue within 2 arcsec
\cite[see e.g.][]{ive02,lu07}, resulting in a sample of 516 quasars.
These 516 quasars were further cross-correlated between the SDSS
quasar positions and the GB6 catalogue within 1 arcmin
\cite[e.g.][]{kim08}. This results in a sample of 212 quasars. The
187 quasars are thus defined as FSRQs conventionally with a spectral
index between 1.4 and 4.85 GHz $\alpha<0.5$. After excluding two
FSRQs due to the weakness of emission lines, our final sample
consists of 185 FSRQs. The source redshift ranges from $\sim0.1$ to
$\sim4.0$, however only one source (SDSS J081009.9+384756.9,
$z=3.95$) has a redshift of $z>3.0$. In 79 out of 185 source, the
redshift is $z<0.8$, enabling us to measure the broad $\rm H\beta$
line, which is commonly used to estimate the black hole mass
\cite[e.g.][]{kas00,gu01}. Based on the identified infrared
counterparts labelled in SDSS DR3 catalogue, the NIR ($J, H, K_{s}$)
data are archived from the Two Micron All Sky Survey \cite[2MASS,
][]{skr06} for 98 sources among 185 sources. These sources were
consistently re-identified through cross-correlating between 2MASS
and SDSS within a matched radius of 2 arcsec, which is much larger
than the sub-arcsec positional accuracy of the SDSS and the 2MASS
surveys. Moreover, we collected the Far- and near-UV magnitudes from
the $Galaxy~ Evolution~ Explorer$ \cite[$GALEX$;][]{mar05} Data
Release 4 matched within 3 arcsec of SDSS positions for 117/185, and
146/185 sources, respectively.

Our sample is listed in Table \ref{tbl-1} and Table \ref{tb2-2}, of
which: (1) Source SDSS name; (2) redshift; (3) black hole mass ($\S$
3.2); (4) synchrotron peak luminosity ($\S$ 4.4); (5) broad line
region luminosity ($\S$ 3.2); (6) bolometric luminosity ($\S$ 3.2);
(7) synchrotron peak frequency ($\S$ 4.4); (8) - (10) spectral
indices between the rest-frame frequencies of 5 GHz, 5000 $\rm \AA$
and 1 keV ($\S$ 5.1).
%\textbf{Table \ref{tb2-2} show the thermal dominated sources }

\section{Parameters derivation}

\subsection{Spectral analysis}
%The spectra of quasars from UV to optical wavelengths are remarkably
%similar to each other.
The spectra of quasars are characterized by a featureless continuum
and various of broad and narrow emission lines \cite[][]{vanden01}.
For our FSRQs sample, we ignore the host galaxy contribution to the
spectrum, since only very little, if any, starlight is observed. In
a first step, the SDSS spectra were corrected for the Galactic
extinction using the reddening map of \cite{sch98} and then shifted
to their rest wavelength, adopting the redshift from the header of
each SDSS spectrum. %Once these steps were completed, we subtracted
%the Fe II complexes, and performed the spectral analysis for the
%emission lines following the procedures described below.
In order to reliably measure line parameters, we choose those
wavelength ranges as pseudo-continua, which are not affected by
prominent emission lines, and then decompose the spectra into the
following three components:

1. A power-law continuum to describe the emission from the active
nucleus. The 15 line-free spectral regions were firstly selected
from SDSS spectra covering 1140$\rm \AA$ to 7180$\rm \AA$ for our
sample, namely, 1140$\rm \AA$ -- 1150$\rm \AA$, 1275$\rm \AA$ --
1280$\rm \AA$, 1320$\rm \AA$ -- 1330$\rm \AA$, 1455$\rm \AA$ --
1470$\rm \AA$, 1690$\rm \AA$ -- 1700$\rm \AA$, 2160$\rm \AA$ --
2180$\rm \AA$, 2225$\rm \AA$ -- 2250$\rm \AA$, 3010$\rm \AA$ --
3040$\rm \AA$, 3240$\rm \AA$ -- 3270$\rm \AA$, 3790$\rm \AA$ --
3810$\rm \AA$, 4210$\rm \AA$ -- 4230$\rm \AA$, 5080$\rm \AA$ --
5100$\rm \AA$, 5600$\rm \AA$ -- 5630$\rm \AA$, 5970$\rm \AA$ --
6000$\rm \AA$, 7160$\rm \AA$ -- 7180$\rm \AA$
\citep{vanden01,Forster01}. Depending on the source redshift, the
spectrum of individual quasar only covers five to eight spectral
regions, from which the initial power-law are obtained for each
source. We found that a single power-law can not give a satisfied
fit for some low-redshift sources whose spectra cover $\rm \Halpha$
and $\rm \Hbeta$ region. In these cases, a double power-law was
adopted to obtain the initial power-law continuum \citep{vanden01}
with the break point around $5100\rm \AA$. %We
%would get better double power law than the initial double power law,
%then using the better overlap the initial to fit again and again,
%because the better the initial data given to fit is, the smaller
%$\rm \chi^2$ of the fitting result is.

2. An Fe II template. The spectra of our sample covers UV and
optical regions, therefore, we adopt the UV Fe II template from
\cite{vestergaard01}, and optical one from V\'{e}ron-Cetty et al.
(2004). The Fe II template obtained by V\'{e}ron-Cetty et al. (2004)
covers the wavelengths between 3535 and 7534 $\rm \AA$, extending
farther to both the blue and red wavelength ranges than the Fe II
template used in \cite{boroson92}. This makes it more advantageous
in modeling the Fe II emission in the SDSS spectra. For the sources
with both UV Fe II and optical Fe II lines prominent in the spectra,
we connect the UV and optical templates into one template covering
the whole spectra \cite[see also][]{Hu08}. %Due to the limitation of
%Fe II template spectral region, only the fitted Fe II line emission
%can be subtracted from the source spectra in the overlapped spectral
%region of Fe II templates and source spectra.
In the fitting, we assume that Fe II has the same profile as the
relevant broad lines, i.e. the Fe II line width usually was fixed to
the line width of broad $\Hbeta$ or $\Mgii$ or $\Civ$, which in most
cases gave a satisfied fit.
%We assume
%that Fe II has the same profile as the broad component of $\rm
%H\beta$.
In some special cases, a free-varying line width was adopted in the
fitting to get better fits.

3. A Balmer continuum generated in the same way as Dietrich et al.
(2002) \cite[see also][]{Hu08}. \cite{grandi82} and \cite{die02}
proposed that a partially optically thick cloud with a uniform
temperature could produce the Balmer continuum, which can be
expressed as:
\begin{equation}
F_{\lambda}^{\rm BaC}=F_{\rm
BE}B_{\lambda}(T_{e})(1-e^{-\tau_{\lambda}}); \qquad (\lambda<
\lambda_{\rm BE}) \label{Flamd.eq}
\end{equation}
where $F_{\rm BE}$ is a normalized coefficient for the flux at the
Balmer edge $(\lambda_{\rm BE}=3646\rm \AA)$, $B_{\lambda}(T_{e})$
is
the Planck function at an electron temperature $T_e$, and %, and
%$B_{\lambda}(T_{e})$ can be expressed by:
%\begin{equation}
%B_{\lambda}(T_{e})=\frac{2hc^2 \lambda^{-5}}{e^{hc/k\lambda T}-1}
%\label{Blamd.eq}
%\end{equation}
%In Equation \ref{Flamd.eq},
$\tau_{\lambda}$ is the optical depth at
 $\lambda$ and is expressed as:
\begin{equation}
\tau_{\lambda}=\tau_{\rm BE}(\frac{\lambda}{\lambda_{\rm BE}})
\label{taulamd.eq}
\end{equation}
where $\tau_{\rm BE}$ is the optical depth at the Balmer edge. There
are two free parameters, $F_{\rm BE}$ and $\tau_{\rm BE}$. Following
\cite{die02}, we adopt the electron temperature to be $T_e \, =\,
15,000$ K.
%At wavelength $\lambda>\lambda_{\rm BE}$, a smooth rise
%from $\sim$ 4000 $\rm \AA$ to the Balmer edge is present due to the
%blended higher order Balmer lines \citep{wills85}. \textbf{However,
%this is ignored in the fitting process due to ... (see Fig.
%\ref{fig2}).}

\begin{figure*}
\includegraphics[width=15cm]{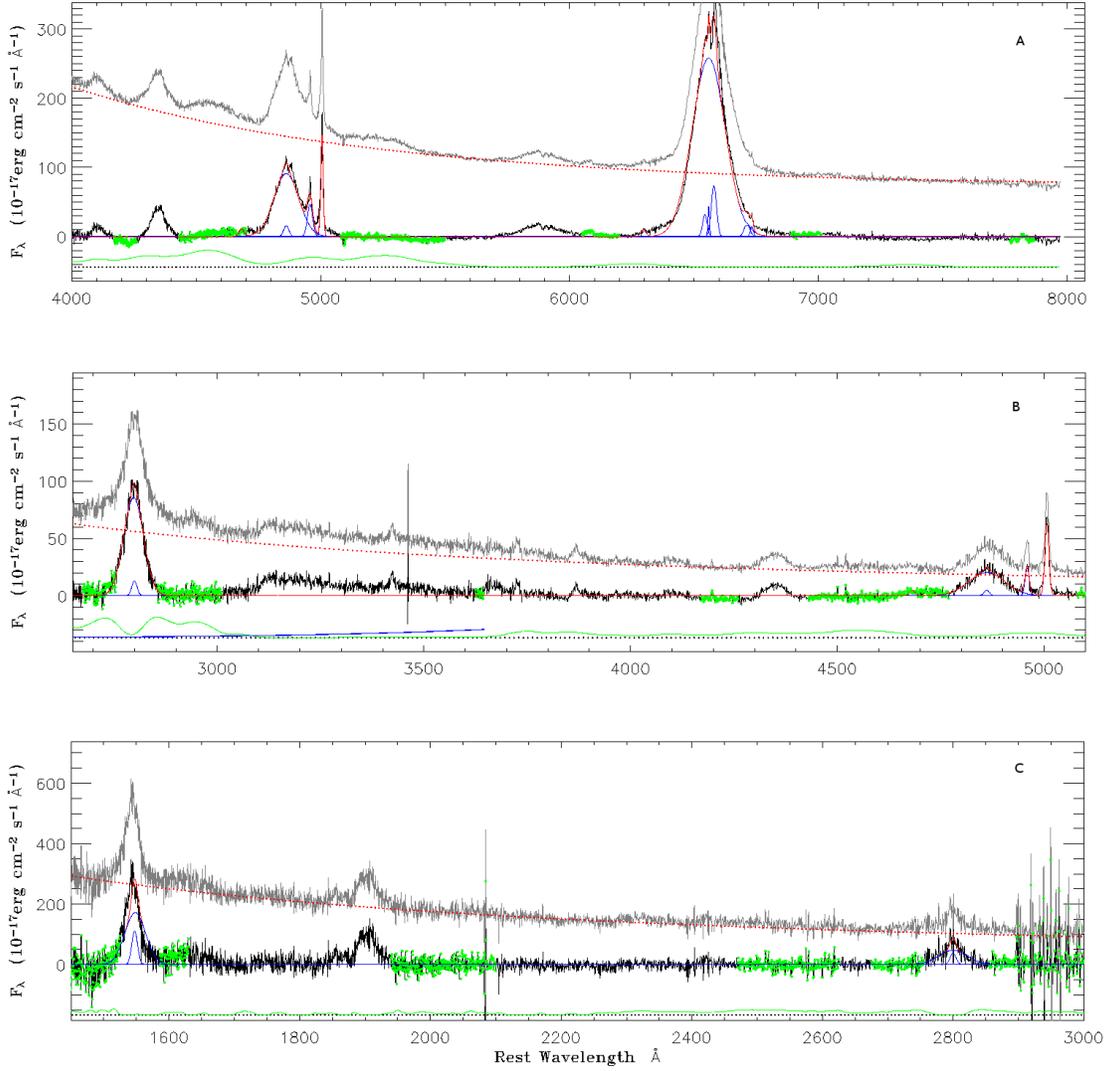}
\caption{The examples of spectral analysis: (A) low redshift FSRQ
SDSS J080131.9+473616.1 ($z=0.157$); (B) middle redshift FSRQ SDSS
J074559.3+331334.5 ($z=0.610$); (C) high redshift FSRQ SDSS
J081303.8+254211.1 ($z=2.024$). In each panel, the top and middle
black lines are the original and the continuum-subtracted spectrum,
respectively. The top dotted red line represents the power-law
continuum. The middle blue lines are the individual line components
in multi-line spectral fitting, and the middle solid red line is
integrated lines fitting. The green spectral region in the
continuum-subtracted spectrum is used to fit Fe II emission and
power-law continuum, and the bottom green line is Fe II emission,
which is shifted downwards with arbitrary unit for the sake of
presentation. The Balmer continuum is indicated as solid blue lines
in the bottom of panel (B), which began at the Balmer edge 3646$\rm
\AA$, however, it is not included in the case of (A) and (C) due to
the wavelength coverage (see text for details).} \label{fig1}
\end{figure*}

The modeling of above three components is performed by minimizing
the $\chi^2$ in the fitting process. The final multicomponent fit is
then subtracted from the observed spectrum. The examples of the
fitted power-law, Fe II lines, Balmer continuum and the residual
spectra for the sources with low, middle and high redshift are shown
in Fig. \ref{fig1}. The Fe II fitting windows are selected as the
regions with prominent Fe II line emission while no other strong
emission lines, according to \cite{vestergaard01} and \cite{kim06}.
The fitting window around the Balmer edge ($3625 - 3645\rm \AA$) is
used to measure the contribution of Balmer continuum, which extends
to $\Mgii$ line region. The Balmer continuum is not considered when
$3625 - 3645\rm \AA$ is out of the spectrum. Therefore, the
inclusion of Balmer continuum depends on the source redshift, which
is illustrated in Fig. \ref{fig1}. While the Balmer continuum should
be included in the fitting for middle redshift sources, it can be
ignored for low and high redshift sources.
%The Balmer continuum only effect the $\Mgii$ line region.
%The selected windows are in the range of $1400 - 8000
%\AA$: 1425$\rm \AA$--1525, 1585--1615, 1600--1630, 1940--2100,
%2470--2625, 2675--2755, 3625--3645, 4170--4260, 4430--4770,
%5080--5550, 6050--6200, 6890--7010, 7790--7890$\rm \AA$.
%see also Hu et al. 2008. The SDSS spectra were firstly transferred
%into source rest frame, and the Galactic extinction was also
%corrected according to ???. To measure the broad emission lines, we
%firstly fit the continuum, which includes a power-law continuum, Fe
%lines and Balmer continuum. In the fitting, the initial power-law
%continuum was fitted from 15 spectral line-free windows in 1140$\rm
%\AA$ to 7180$\rm \AA$ (Vanden01) and Forster01. In case of Fe lines,
%we used optical template of Boroson \& Green (1992) and UV Fe
%template from Vestergaard \& Wilkes (2001), and the Fe templates
%were broadened to match the source spectra during fitting. The final
%fitting was performed through $\chi^{2}$ minimization method, and
%then was subtracted from the spectra.

The broad emission lines were measured from the continuum subtracted
spectra. We mainly focused on several prominent emission lines, i.e.
${\rm Ly\,\alpha}, \Halpha, \Hbeta$, Mg II, C IV. Generally, two
gaussian components were adopted to fit each of these lines,
indicating the broad and narrow line components, respectively. The
blended narrow lines, e.g. [O III] $\lambda\lambda4959,5007\rm \AA$
and [He II] $\rm \lambda 4686 \AA$ blending with $\rm H\beta$, and
[S II] $\lambda\lambda6716,6730\rm \AA$, [N II]
$\lambda\lambda6548,6583\rm \AA$ and [O I]$\rm \lambda 6300 \AA$
blending with $\rm H\alpha$, were included as one gaussian component
for each line at the fixed line wavelength. The $\chi^{2}$
minimization method was used in fits. The line width FWHM, line flux
of broad ${\rm Ly\,\alpha}, \Halpha, \Hbeta$, Mg II and C IV lines
were obtained from the final fits for our sample. The examples of
the fitting are shown in Fig. \ref{fig1} for the sources in the
different redshift.

\subsection{$M_{\rm bh}$ and $L_{\rm BLR}$} \label{mbhlbol}

There are various empirical relations between the radius of broad
line region (BLR) and the continuum luminosity, which can be used to
calculate the black hole mass in combination with the line width
FWHM of broad emission lines. However, there are defects when using
the continuum luminosity to estimate the BLR radius for blazars
since the continuum flux of blazars are usually doppler boosted due
to the fact that the relativistic jet is oriented close to the line
of sight. Alternatively, the broad line emission can be a good
indicator of thermal emission from accretion process. Therefore for
our FSRQs sample, we estimate the black hole mass by using the
empirical relation based on the luminosity and FWHM of broad
emission lines. According to the source redshift, we use various
relations to estimate the black hole mass: Greene et al. (2005)
relation for broad $\rm H\alpha$ line; Vestergaard et al. (2006) for
broad $\rm H\beta$ and Kong et al. (2006) for broad Mg II and C IV
lines.

\cite{Green05} provided a formula to estimate the black hole mass
using the line width and luminosity of broad $\Halpha$ alone, which
is expressed as
%%%% second selected
%%%%%%%%%%%%%%%%%%%%%%%%%%%%%%%%%%%%%%%%%%%%%%%%%%%%%%%%%%%%%%%%%%%
\begin{equation}
%\begin{array}{l}
M_{\rm BH}(\Halpha)=(2.0^{+0.4}_{-0.3})\times 10^6\Big(\frac{{\it
L}(\Halpha)}{10^{42}\,erg\,s^{-1}}\Big)^{0.55\pm0.02}~
\Big(\frac{{\rm FWHM}(\Halpha)}{10^3\,km\,s^{-1}}\Big)^{2.06\pm0.06}
{\it M}_{\sun} \label{MBH_LHa.eq}
% \end{array}
\end{equation}
%%%%%%%%%%%%%%%%%%%%%%%%%%%%%%%%%%%%%%%%%%%%%%%%%%%%%%%%%%%%%%%%%%%

For the sources with available FWHM and luminosity of broad
$\Hbeta$, the method to calculate $M_{\rm BH}$ is given by
\cite{Vestergaard06}:
%%%%%%%%%first selected
%%%%%%%%%%%%%%%%%%%%%%%%%%%%%%%%%%%%%%%%%%%%%%%%%%%%%%%%%%%%%%%%%%%
\begin{equation}
%\begin{array}{c}
M_{\rm BH} (\Hbeta) = 4.68\times10^6 % \log \,
   \left( \frac{\it L(\rm H\beta)}{10^{42}\, \rm erg~s^{-1}}\right)^{0.63}
  \left(\frac{\rm FWHM(H\beta)}{1000~km~s^{-1}} \right)^2 {\it M}_{\sun}%\\
%   + (6.67 \pm 0.03).
\label{MBH_LHb.eq}
%\end{array}
\end{equation}
%%%%%%%%%%%%%%%%%%%%%%%%%%%%%%%%%%%%%%%%%%%%%%%%%%%%%%%%%%%%%%%%%%%

In addition, \cite{kong06} presented the empirical formula to obtain
the black hole mass using broad $\Mgii$ and $\Civ$ for high redshift
sources as follows,
    %%% average these two Mbh, if have two Mbh
%%%%%%%%%%%%%%%%%%%%%%%%%%%%%%%%%%%%%%%%%%%%%%%%%%%%%%%%%%%%%%%%%%%
\begin{equation}
  M_{\rm BH}({\rm Mg~II}) = 2.9 \times 10^{6} \left
  (\frac{L({\rm Mg~II})}{\rm 10 ^{42}\, erg\, s^{-1}}\right )^{0.57\pm0.12}
  \left(\frac{\rm FWHM(Mg~II)}{\rm 1000\, km\, s^{-1}}\right )^{2}
  M_{\sun},
\label{MBH_Mgii.eq}
\end{equation}

\begin{equation}
  M_{\rm BH}({\rm C~IV}) = 4.6 \times 10^{5} \left
  (\frac{L({\rm C~IV})}{\rm 10 ^{42}\, erg\, s^{-1}}\right )^{0.60\pm0.16}
  \left(\frac{\rm FWHM(C~IV)}{\rm 1000\, km\, s^{-1}}\right )^{2} M_{\sun}
\label{MBH_Civ.eq}
\end{equation}
%%%%%%%%%%%%%%%%%%%%%%%%%%%%%%%%%%%%%%%%%%%%%%%%%%%%%%%%%%%%%%%%%%%

In the redshift range of our sample, $M_{\rm BH}$ can be estimated
using two of above relations for 122 out of 185 FSRQs. In most
sources (113/122), two $M_{\rm BH}$ values are consistent with each
other within a factor of three. For low-redshift sources, we first
selected the black hole mass estimated from broad $\Hbeta$ line,
which is commonly used to estimate the black hole mass for
low-redshift sources \cite[e.g.][]{kas00,gu01}. In the case that the
spectral quality or the spectral fitting of $\Hbeta$ region is poor,
we instead use broad $\Halpha$ line to estimate the black hole mass.
Moreover, we adopted the average value of $M_{\rm BH}~ ({\rm
Mg~II})$ and $M_{\rm BH}~ ({\rm C~IV})$ when both values are
available for one source.

In this work, the BLR luminosity $L_{\rm BLR}$ is derived following
\cite{cel97} by scaling the strong broad emission lines ${\rm
Ly\,\alpha}, \Halpha, \Hbeta$, Mg II and C IV to the quasar template
spectrum of Francis et al. (1991), in which Ly$\alpha$ is used as a
reference of 100. By adding the contribution of $\rm H\alpha$ with a
value of 77, the total relative BLR flux is 555.77, of which $\rm
Ly\alpha$ is 100, $\Halpha$ 77, $\Hbeta$ 22, $\Mgii$ 34, and $\Civ$
63 \citep{cel97,Francis91}. From the BLR luminosity, we estimate the
bolometric luminosity as $L_{\rm bol}=10L_{\rm BLR}$
\citep{Netzer90}.

%Broad Line Region total Luminosity($\rm L_{BLR}$) can be inferred by
%\cite{cel97,Francis91}
%\begin{equation}
%L_{\rm BLR}=\frac{555\times(L_{\rm
%Ly\,\alpha}+L_{\Halpha}+L_{\Hbeta}+L_{\Mgii}+L_{\Civ})}{(100+77+22+34+63)}
%\label{L_BLR.eq}
%\end{equation}
%Where the parameters of  $\rm Ly\alpha \,\lambda\,$1216 is 100,
%$\Halpha \,\lambda\,$ 6562 is 77, $\Hbeta\,\lambda\,$4861 is 22,
%$\Mgii\,\lambda\,$2800 is 34, $\Civ \, \lambda\,$1549 is 63.

\section{The thermal emission} \label{refine}

In addition to the relativistically beamed, non-thermal jet
emission, the thermal emission from the accretion disk and the host
galaxy are expected to be present in radio quasars.  In some cases,
the thermal emission can be dominated over the nonthermal jet
emission \cite[e.g.][]{landt08}. We estimated the contribution of
thermal emission in SEDs as follows \cite[see also][]{landt08}, and
our sample is thus refined to include only the sources with SEDs
dominated by the nonthermal jet emission.

\subsection{The accretion disk}

   Following \cite{del03} and \cite{landt08}, we calculated accretion
disk spectra assuming a steady geometrically thin, optically thick
accretion disk. In this case the emitted flux is independent of
viscosity, and each element of the disk face radiates roughly as a
blackbody with a characteristic temperature, which depends only on
the mass of the black hole, $M_{\rm BH}$, the accretion rate,
$\dot{M}$, and the radius of the innermost stable orbit (e.g.,
Peterson 1997; Frank et al. 2002). We have adopted the Schwarzschild
geometry (nonrotating black hole), and for this the innermost stable
orbit is at $r_{\rm in}=6r_{\rm g}$, where $r_{\rm g}$ is the
gravitational radius defined as $r_{\rm g}=GM_{\rm BH}/c^{2}$, G is
the gravitational constant, and c is the speed of light.
Furthermore, we have assumed that the disk is viewed face-on. The
accretion disk spectrum is fully constrained by the two quantities,
accretion rate and mass of the black hole. We have calculated the
accretion rate using the relations $L_{\rm bol}=
\epsilon\dot{M}c^{2}$, where $\epsilon$ is the efficiency for
converting matter to energy, with $\epsilon\sim6\%$ in the case of a
Schwarzschild black hole. The bolometric luminosity is estimated as
$L_{\rm bol}=f^{-1}L_{\rm BLR}$ with $f$ the BLR covering factor,
which is not well known, and can be in the range of $\sim5\%-30\%$
(Maiolino et al. 2001 and references therein). As in Section
\ref{mbhlbol}, we adopt a canonical value of $f\sim10\%$ (Peterson
1997). The contribution of accretion disk thermal emission are
estimated by calculating the fraction of the thermal emission to the
SED data at SDSS optical and $GALEX$ UV region. Tentatively, we
simply use a marginal value of 50\% at most of SED wavebands to
divide the FSRQs into thermal-dominated ($>50\%$) and
nonthermal-dominated ($<50\%$), and found that the thermal emission
can be dominant in 100/185, 35/185, 2/185 FSRQs for $f=$ 5\%, 10\%,
and 30\%, respectively.

\subsection{The host galaxy}

%\textbf{To evaluate the contribution of host galaxy thermal emission
%to the optical spectrum, especially for low-redshift (e.g. $z<0.3$)
%sources, we directly investigate the SDSS spectra by estimating Ca
%H\&K break, which is commonly used to classify the blazars and radio
%galaxies \citep{mar96,lan02,lan04,col05}. The Ca H\&K break is a
%prominent absorption feature typically seen in the spectra of
%elliptical galaxies (the hosts of radio-loud AGN; e.g. Wurtz, Stocke
%\& Yee 1996) and is located at $\rm \sim4000\AA$ rest frame
%wavelength. It is defined as $C=(f_{+}-f_{-})/f_{+}$, where $f_{-}$
%and $f_{+}$ are the fluxes in the rest frame wavelength regions $\rm
%3750-3950\AA$ and $\rm 4050-4250\AA$ respectively. Its value in
%normal nonactive ellipticals is on average $\sim0.5$ (Dressler \&
%Shectman 1987). In blazars the value of the Ca H\&K break is assumed
%to be decreased by the presence of non-thermal jet emission
%\citep{lan04}.}

   Usually, the host galaxies of radio quasars are bright ellipticals,
and their luminosity only spread in a relatively narrow range
\cite[e.g.][]{mcl04}. We estimated the contribution from host galaxy
thermal emission using the elliptical galaxy template of
\cite{man01}, which extends from near-IR to UV frequencies \cite[see
also][]{landt08}. Slightly different from \cite{landt08}, we use the
bulge absolute luminosity in R-band estimated from $M_{\rm BH} -
M_{\rm R}$ relation of \cite{mcl04} in combination with the estimted
black hole mass in Section \ref{mbhlbol},
\begin{equation}
\rm log~ \it M_{\rm BH}/\rm M_{\odot}=-0.50(\pm0.02)\it M_{\rm R}\rm
-2.74(\pm0.48)
\end{equation}
The contribution of host galaxy thermal emission are estimated by
comparing the calculated thermal emission with the SED data at SDSS
optical, $GALEX$ UV and 2MASS NIR region. The value of 50\% is used
to distinguish thermal-dominated with thermal emission $>50\%$ and
non-thermal dominated with thermal emission $<50\%$. We found that
the thermal emission can be dominant in about 16 of 185 FSRQs.

\subsection{Sample refinement}

  Although the thermal emission may dominate in only small fraction of
sources, i.e. accretion disk thermal emission in 35 sources, and
host galaxies emission in 16 sources, we combine the contribution of
accretion disk and host galaxy to maximize the source number, of
which the non-thermal jet emission is not dominated in SED. We
simply add the expected thermal emission from accretion disk (using
a canonical value $f=10\%$) and host galaxy, then compare with the
SED data at SDSS optical, GALEX UV and 2MASS NIR region. Finally, we
found the combined thermal emission can dominate over ($>50\%$)
nonthermal jet emission in about 67 sources, which are then called
thermal-dominated FSRQs in this work and listed in Table
\ref{tb2-2}. The remaining 118 sources are recognized as nonthermal
jet-dominated FSRQs in this work (see Table \ref{tbl-1}).

\subsection{$\nupeak$ and $\nuLupeak$} \label{nln}

\begin{figure}
\center
\includegraphics[width=15.0cm]{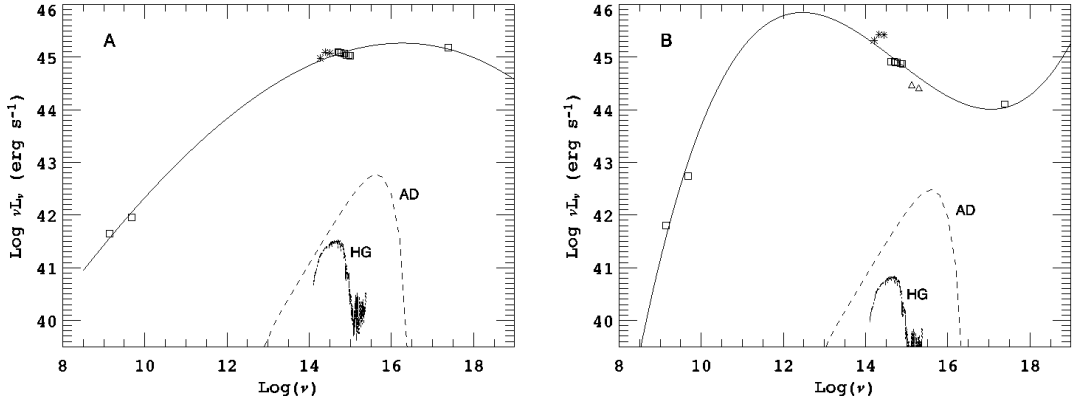}\\
\caption{Representative three-degree polynomial fits to the spectral
energy distributions of FSRQs in our sample. Open rectangles are
data from FIRST, GB6, SDSS and RASS. Stars represent data from
2MASS, and triangles for $GALEX$ UV data. The solid line is the SED
fit, while the dotted one (HG) is the calculated emission from host
galaxy. The dashed line (AD) is the expected emission from the
accretion disk at assuming $f=10\%$ (see text for details). The
synchrotron peak frequencies for the two sources are
$\nupeak=1.86\times10^{16}$ Hz (SDSS J150324.7+475830.4, $z=0.345$;
A) and $\nupeak=2.88\times10^{12}$ Hz (SDSS J083148.9+042939.2,
$z=0.174$; B). See text for details.} \label{SEDsample}
\end{figure}

The SED of each quasar was constructed from multi-band data, which
covers radio (1.4 and 4.85 GHz), optical (5 - 8 line-free windows
selected from SDSS spectra), and X-ray (1keV) data. The optical
continuum were picked out with five to eight line-free regions
\citep{Forster01,vanden01} from the SDSS spectra in the source rest
frame, of which no or only very weak Fe lines are present (see
Section 3.1). The radio flux at 1.4 and 4.85 GHz are K-corrected to
the source rest frame using the spectral index between these two
frequencies. The X-ray 1 keV luminosity are obtained after
K-correction and correcting the Galactic extinction \citep{Shen06}.
The IR ($J$, $H$, and $K_{s}$) data collected from 2MASS
\citep{skr06} were also added to construct SEDs, which are available
for 98 FSRQs. Moreover, the Far- (for 117 sources) and near-UV (for
146 sources) data are also added in constructing SEDs, after
correcting the Galactic extinction and K-correction using a spectral
index of 0.5. Considering the possibility that the X-ray emission of
FSRQs can be from the inverse Compton process, we fitted the data
points for each source (in a $\rm \nu$ versus $\rm \nu L_\nu$
diagram) with a third-degree polynomial following \cite{fossati98},
which yields an upturn allowing for X-ray data-points that do not
lie on the direct extrapolation from the lower energy spectrum (see
examples in Fig. \ref{SEDsample}). Through fitting, we obtained the
synchrotron peak frequency and the corresponding peak luminosity for
each source. In following analysis, we will present the analysis
only for the non-thermal jet-dominated FSRQs (see Table
\ref{tbl-1}).

\section{results}

\subsection{$\alpha_{\rm ro}$ -- $\alpha_{\rm ox}$ plane}

\begin{figure}
\center
\includegraphics[width=8.5cm]{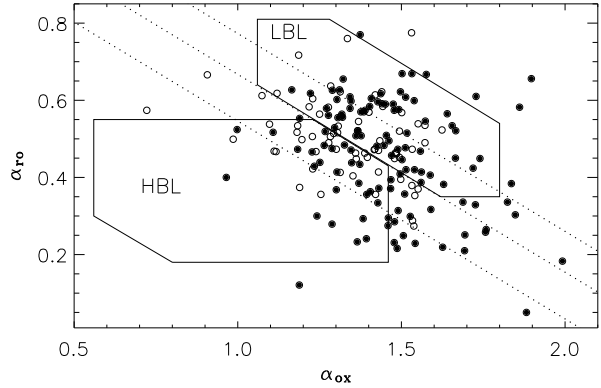}\\
\caption{The $\alpha_{\rm ro}$ \& $\alpha_{\rm ox}$ relation. The
effective spectral indices are defined in the usual way and
calculated between the rest-frame frequencies of 5 GHz, $\rm
5000~\AA$, and 1 keV. The solid circles are 118 nonthermal-dominated
FSRQs, while the open ones for 67 thermal-dominated sources (see
text for details).
%The dashed line represents the locus of
%constant $\alpha_{\rm rx}=0.78$, with $\alpha_{\rm rx}\gtrsim0.78$
%and $\alpha_{\rm rx}\lesssim0.78$ being typical of BL Lacs with
%X-rays dominated by inverse Compton (LBL) and synchrotron radiation
%(HBL), respectively. The ``HBL box'' as defined by \cite{padovani03}
%represents the $2\sigma$ region around the mean $\alpha_{\rm ro}$,
%$\alpha_{\rm ox}$, and $\alpha_{\rm rx}$ values of HBL.
The dotted lines represent, from top to bottom, the loci of
$\alpha_{\rm rx}=0.85$, typical of 1 Jy FSRQs and LBLs; $\alpha_{\rm
rx}=0.78$, the dividing line between HBLs and LBLs; and $\alpha_{\rm
rx}=0.70$, typical of RGB BL Lac objects. The `HBL' and `LBL' boxes
defined in Padovani et al. (2003) are indicated by the solid lines
and marked accordingly, which represent the regions within $2\sigma$
from the mean $\alpha_{\rm ro}$, $\alpha_{\rm ox}$, and $\alpha_{\rm
rx}$ values of HBLs and LBLs, respectively (see also Fig. 1 in
Padovani et al. 2003).} \label{rox}
\end{figure}

The broadband properties of our sources can be firstly studied by
deriving their $\alpha_{\rm ox}$, $\alpha_{\rm ro}$, and
$\alpha_{\rm rx}$ values, which are the usual rest-frame effective
spectral indices defined between 5 GHz, 5000 $\rm \AA$, and 1 keV.
The 5000 $\rm \AA$ optical continuum flux density is derived (or
extrapolated when 5000 $\rm \AA$ is out of the SDSS spectral region)
from the direct power-law fit on the 5 - 8 selected line-free SDSS
spectral windows in the source rest frame (see Section 3.1). The 5
GHz flux density have been k-corrected using the spectral index
between FIRST 1.4 GHz and GB6 4.85 GHz. In Fig. \ref{rox}, we
present $\alpha_{\rm ro}$ - $\alpha_{\rm ox}$ relation for the
sample. Following \cite{padovani03}, three lines are indicated:
$\alpha_{\rm rx}=0.85$, typical of 1 Jy FSRQs and LBLs; $\alpha_{\rm
rx}=0.78$, the dividing line between HBLs and LBLs; and $\alpha_{\rm
rx}=0.70$, typical of RGB BL Lac objects. Moreover, the `HBL' and
`LBL' boxes defined in \cite{padovani03} are also indicated in the
figure, which represent the regions within $2\sigma$ from the mean
$\alpha_{\rm ro}$, $\alpha_{\rm ox}$, and $\alpha_{\rm rx}$ values
of HBLs and LBLs in the multifrequency AGN catalog of \cite{pad97},
respectively (see Fig. 1 in Padovani et al. 2003). This ``HBL box''
is expected to be populated by high-energy peaked blazars, both BL
Lacs and FSRQs. Among total 118 FSRQs, 48 sources have $\alpha_{\rm
rx}<0.78$, of which 28 sources locate in HBL box. In contrast, 59
sources among 70 $\alpha_{\rm rx}>0.78$ sources are in LBL box.

\subsection{$\nupeak$ and $\nuLupeak$}

\begin{figure}
\center
\includegraphics[width=8.5cm]{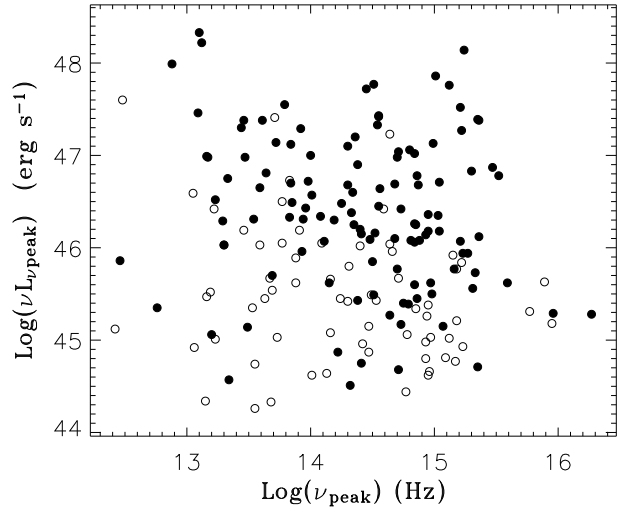}\\
\caption{The synchrotron peak frequency and the peak luminosity for
our sample. The symbols are same as in Fig. \ref{rox}.}
\label{nuLupeak.ps}
\end{figure}

\begin{figure}
\center
\includegraphics[width=8.5cm]{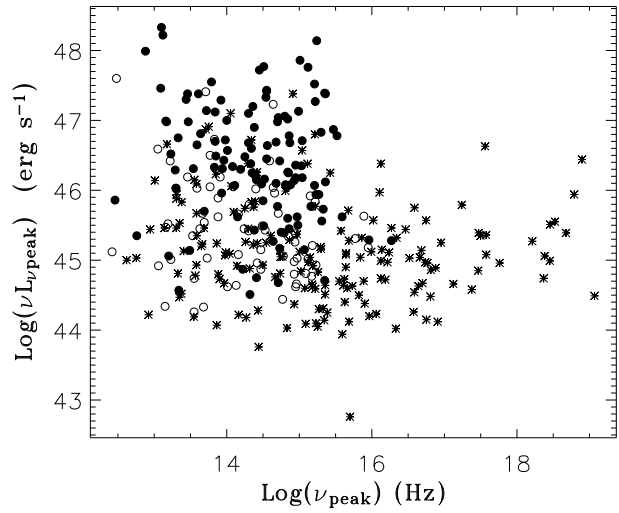}\\
\caption{The synchrotron peak frequency and the peak luminosity. The
circles are same as in Fig. \ref{rox}, while the stars are 170 BL
Lac objects from \citet{wu07}.} \label{allnuLup.ps}
\end{figure}

The relation of the synchrotron peak frequency and peak luminosity
is presented in Fig. \ref{nuLupeak.ps}. We found only a weak
anti-correlation between the synchrotron peak frequency $\nupeak$
and the peak luminosity $\nuLupeak$ with the Spearman correlation
coefficient $r=-0.161$ at $\sim 92\%$ confidence level. The
$\nupeak$ distribution ranges $10^{12.4}$ and $10^{16.3}$ Hz for
whole sample, and between $10^{13}$ and $10^{15.5}$ Hz for most
(111/118) of sources, with $\langle \rm log~
\nupeak\rangle=14.41\pm0.74$ Hz for whole sample. We found
$\nupeak>10^{15}$ Hz in 25 sources (see Fig. \ref{nuLupeak.ps}),
which is larger than the typical value of FSRQs \citep{fossati98}.
As outliers to blazar sequence, the blue quasars are supposed to
have large $\nupeak$ and high $\nuLupeak$ as
well \cite[see e.g.][]{padovani03}. %Differently, the peak luminosity
%of these 25 sources are relatively low though the peak frequency is
%quite large $\nupeak>10^{15}$ Hz.
To further understand the nature of these high $\nupeak$ FSRQs, we
combine our sample with the sample of 170 BL Lac objects in
\cite{wu07}. A significant anti-correlation between $\nupeak$ and
$\nuLupeak$ is present with the Spearman correlation coefficient
$r=-0.343$ at $\gg$ 99.99\% confidence level for the combined sample
of 288 blazars, which is shown in Fig. \ref{allnuLup.ps}. This
anti-correlation covers about seven order of magnitude in $\nupeak$
for the combined sample. Our FSRQs with high $\nupeak>10^{15}$ Hz
have systematically higher peak luminosity than that of BL Lacs,
with $\langle \rm log~ \nuLupeak\rangle=46.41\pm0.94$ compared to
$44.90\pm0.71$ for BL Lacs at same $\nupeak$ range, and the peak
luminosity of some FSRQs are out of the range covered by BL Lacs.
However, we found that the thermal emission from accretion disk can
be dominated over nonthermal jet emission in 11 of 25 sources with
$\nupeak>10^{15}$ Hz if a BLR covering factor of $5\%$ is assumed.
Therefore, the possibility that the contribution from the thermal
emission causes the high synchrotron peak frequency can not be
completely excluded. Although high $\nupeak$ are found in our FSRQs,
they do not reach the extreme value of HBLs, which is also found in
DXRBS sample when comparing high $\nupeak$ FSRQs with BL Lacs
\citep{padovani03}. At the lower-left corner of Fig.
\ref{allnuLup.ps}, there are some FSRQs with relatively low
synchrotron peak frequency as well as low luminosity, comparable to
some low luminosity LBLs. While only a weak anti-correlation present
between $\nupeak$ and $\nuLupeak$ for \cite{wu07} BL Lacs sample, it
becomes significant when combining with our sample. To some extent,
this reflects the importance of sample selection in investigating
the blazar sequence \cite[e.g.][]{padovani07}. Although the
anti-correlation is significant in Fig. \ref{allnuLup.ps}, we notice
that the scatter is significant, mainly caused by the sources in the
lower-left corner. However, the extreme sources with high $\nupeak$
as well as high peak luminosity (i.e. at upper-right corner) are
still lacking. On the other hand, the large scatter may imply that
other parameters may be at work, and the scatter can be much lower
once this parameter is included.

\begin{figure}
\center
\includegraphics[width=8.5cm]{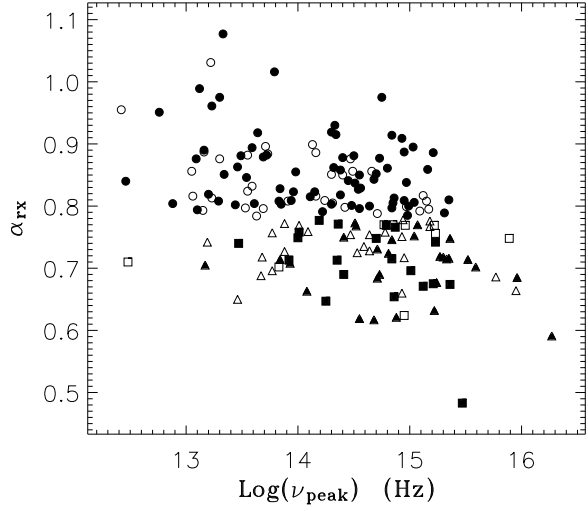}\\
\caption{The synchrotron peak frequency and $\alpha_{\rm rx}$. The
circles represent FSRQs with $\alpha_{\rm rx}>0.78$, the triangles
for FSRQs in HBL box, and the rectangles are FSRQs out of HBL box
however with $\alpha_{\rm rx}<0.78$. The solid symbols are for 118
nonthermal-dominated FSRQs, while the open ones for 67
thermal-dominated sources.} \label{rxnp}
\end{figure}

\begin{figure}
\center
\includegraphics[width=8.5cm]{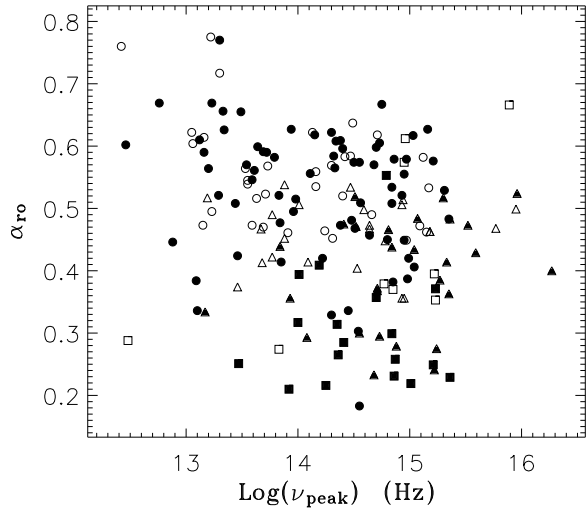}\\
\caption{The synchrotron peak frequency and $\alpha_{\rm ro}$. The
symbols are same as those in Fig. \ref{rxnp}.} \label{ronp}
\end{figure}

We investigate the relationship between $\alpha_{\rm rx}$ and
$\nupeak$ for our sample in Fig. \ref{rxnp}. A significant
anti-correlation is found with a Spearman correlation coefficient
$r=-0.443$ at confidence level $\gg99.99\%$. However, the
considerable scatter is also present with the scatter in $\nupeak$
more than one order of magnitude for any given value of $\alpha_{\rm
rx}$. The mean $\nupeak$ of FSRQs in HBL box is only slightly larger
than that of sources out of the HBL box with a factor of four.
Consequently, the FSRQs in HBL box do not exclusively have high
$\nupeak$. The $\nupeak$ of souces in HBL box are comparable to
those with $\alpha_{\rm rx}<0.78$ but out of HBL box. This actually
can be reflected from the relationship between $\alpha_{\rm ro}$ and
$\nupeak$ shown in Fig. \ref{ronp}. The significant anti-correlation
is found at a confidence level of $\sim99.9\%$, however the scatter
is significant. At any given value of $\alpha_{\rm ro}$, the source
could have a low $\nupeak$ or a high $\nupeak$. As a result, the
sources out of HBL box with $\alpha_{\rm rx}<0.78$ are not necessary
to have lower or higher $\nupeak$ than that of sources in HBL box,
though the latter show a wider $\alpha_{\rm ro}$ range. Consistent
with the anti-correlation in Fig. \ref{rxnp}, the mean $\nupeak$
value of FSRQs with $\alpha_{\rm rx}<0.78$ is larger with a factor
of four than that of $\alpha_{\rm rx}>0.78$ sources. Owing to the
large scatters in both $\alpha_{\rm rx} - \nupeak$ and $\alpha_{\rm
ro} - \nupeak$ relations, it seems that neither solely $\alpha_{\rm
rx}$ nor $\alpha_{\rm rx}$ - $\alpha_{\rm rx}$ combination can
precisely predict the $\nupeak$ value for our present FSRQs sample.

\subsection{($\nupeak$, $\nuLupeak$) \& ($M_{\rm
bh}$, $L_{\rm bol}/L_{\rm Edd}$)}

\begin{figure}
\center
\includegraphics[width=8.5cm]{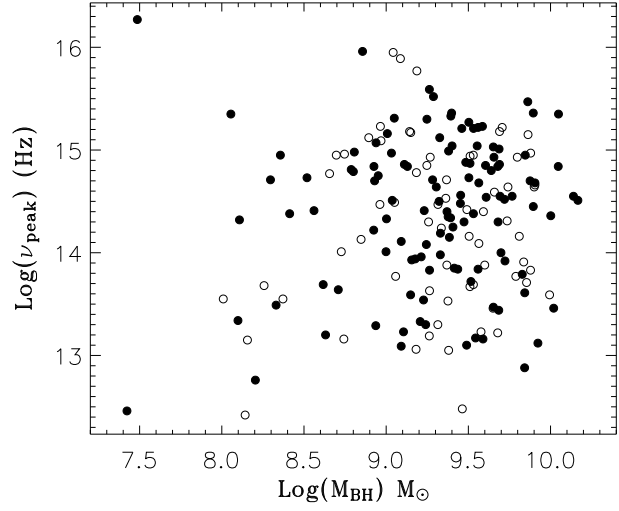}\\
\caption{The synchrotron peak frequency versus the black hole mass.
The symbols are same as in Fig. \ref{rox}.} \label{nuMBH.ps}
\end{figure}

\begin{figure}
\center
\includegraphics[width=8.5cm]{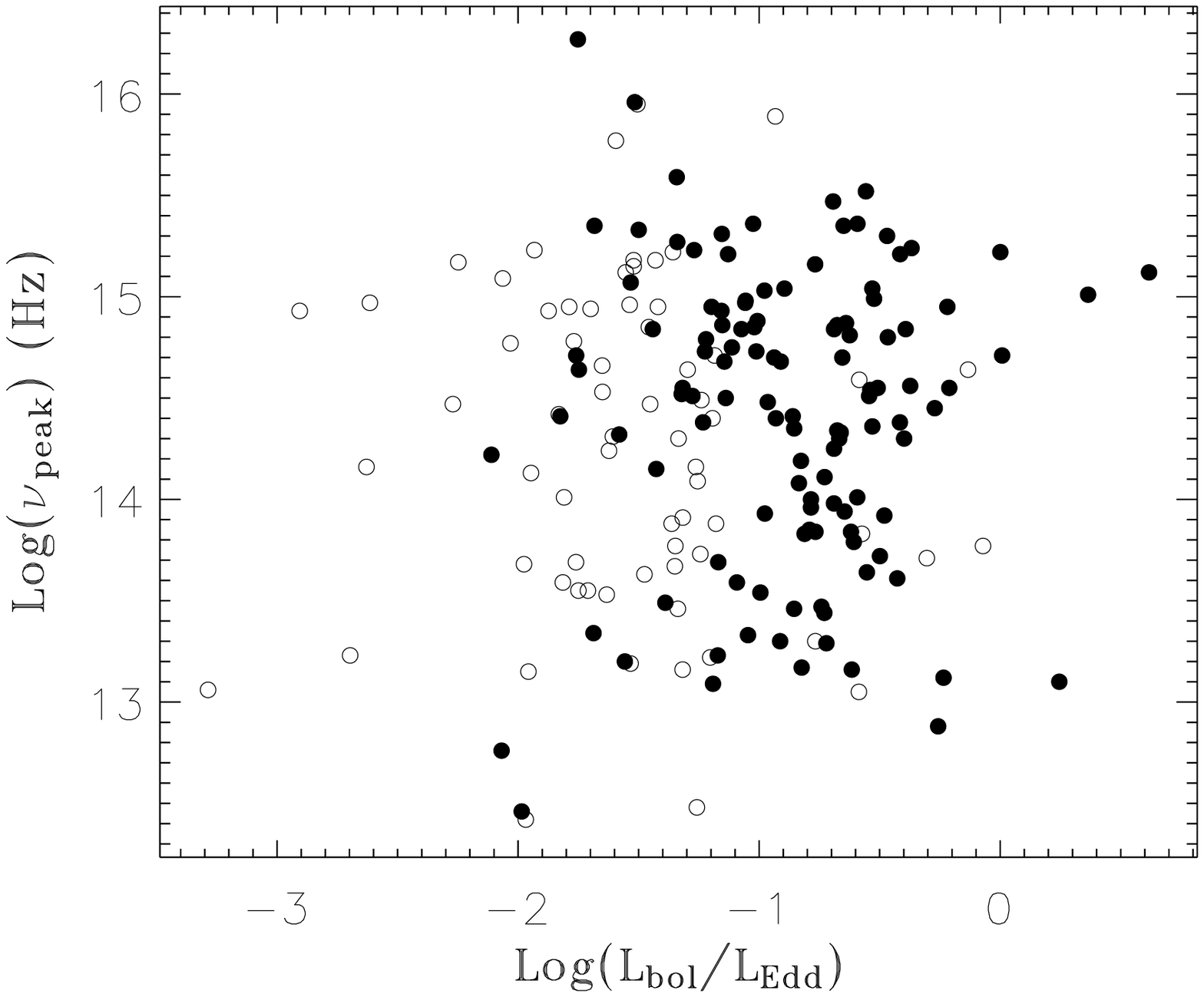}
\caption{The synchrotron peak frequency versus $\Lubol/ \LuEdd$. The
symbols are same as in Fig. \ref{rox}.} \label{nuLulobEdd.ps}
\end{figure}

The estimation of black hole mass and BLR luminosity (then the
corresponding bolometric luminosity) enable us to investigate the
relationship between the synchrotron emission and the accretion
process. In Figs. \ref{nuMBH.ps} and \ref{nuLulobEdd.ps}, we show
the relation of $\nupeak$ \& $M_{\rm bh}$, and $\nupeak$ \& $L_{\rm
bol}/L_{\rm Edd}$, respectively. The black hole mass ranges from
$10^{7.4}$ to $10^{10.5}~ M_{\odot}$ with most of sources
($\sim86\%$; 102 out of 118 sources) in the range of $10^{8.5} -
10^{10}~ M_{\odot}$, while the eddington ratio $L_{\rm bol}/L_{\rm
Edd}$ ranges from $\sim 10^{-2}$ to $\sim 10^{0.6}$, with most of
source ($\sim94\%$; 111 out of 118 sources) in the range of 0.01 to
1. No strong correlations are found either between $\nupeak$ and
$M_{\rm
bh}$, or between $\nupeak$ \& $L_{\rm bol}/L_{\rm Edd}$. % with the
%Spearman correlation coefficient
%$r=0.169$ at confidence level $\sim$ 98\%. %The result indicates that
%the jet physics may be tightly related with the accretion process.

\begin{figure}
\center
\includegraphics[width=8.5cm]{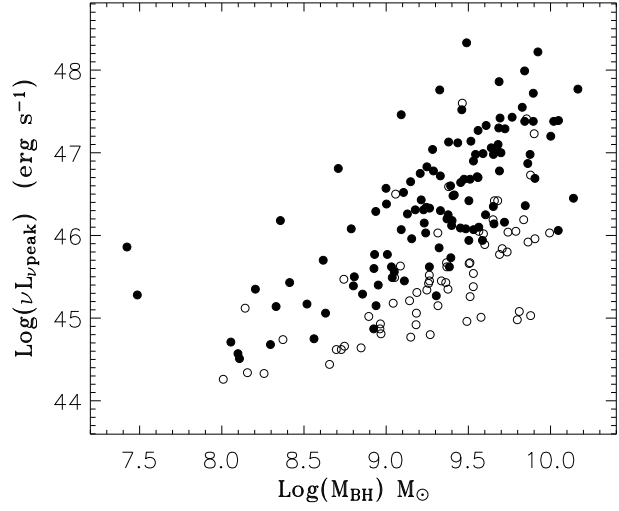}
\caption{The synchrotron peak luminosity versus the black hole mass.
The symbols are same as in Fig. \ref{rox}.} \label{lupeakMBH.ps}
\end{figure}

\begin{figure}
\center
\includegraphics[width=8.5cm]{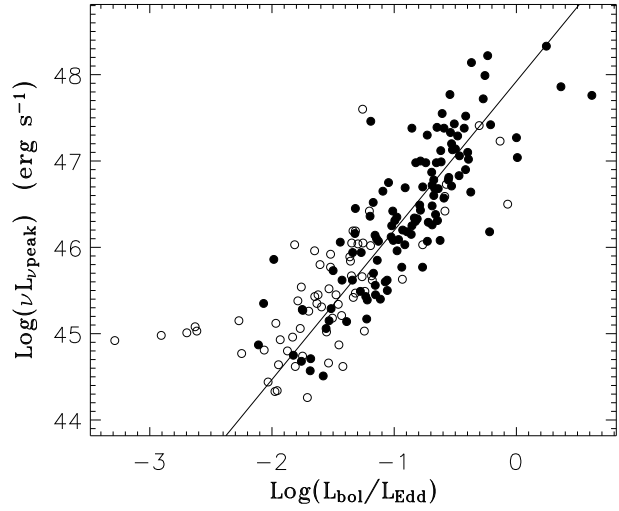}
\caption{The synchrotron peak luminosity versus $\Lubol/ \LuEdd$.
The symbols are same as in Fig. \ref{rox}. The solid line is the OLS
bisector linear fit for 118 nonthermal-dominated FSRQs.}
\label{LulobEddLupeak.ps}
\end{figure}

We found a significant correlation between the black hole mass
$M_{\rm BH}$ and the synchrotron peak luminosity $\nuLupeak$ with
the Spearman correlation coefficient $r=0.724$ at $\sim 99.99\%$
confidence level, which is shown in Fig. \ref{lupeakMBH.ps}.
Moreover, a significant correlation between $\nuLupeak$ and $\Lubol/
\LuEdd$ is also found with the Spearman correlation coefficient
$r=0.842$ at $\gg$ 99.99\% confidence level (see Fig.
\ref{LulobEddLupeak.ps}). Both correlations still present when we
perform the Spearman partial correlation analysis to exclude the
common dependence on redshift. The ordinary least-square (OLS)
bisector linear fit to $\nuLupeak$ and $\Lubol/ \LuEdd$ gives,
\begin{equation}
\rm log~ \it \nuLupeak = \rm (1.73\pm0.11)~ \rm log~ \it \Lubol/
\LuEdd + \rm (47.92\pm0.11)
\end{equation}
This result indicates that the jet physics may be tightly related
with the accretion process.

\begin{figure}
\center
\includegraphics[width=8.5cm]{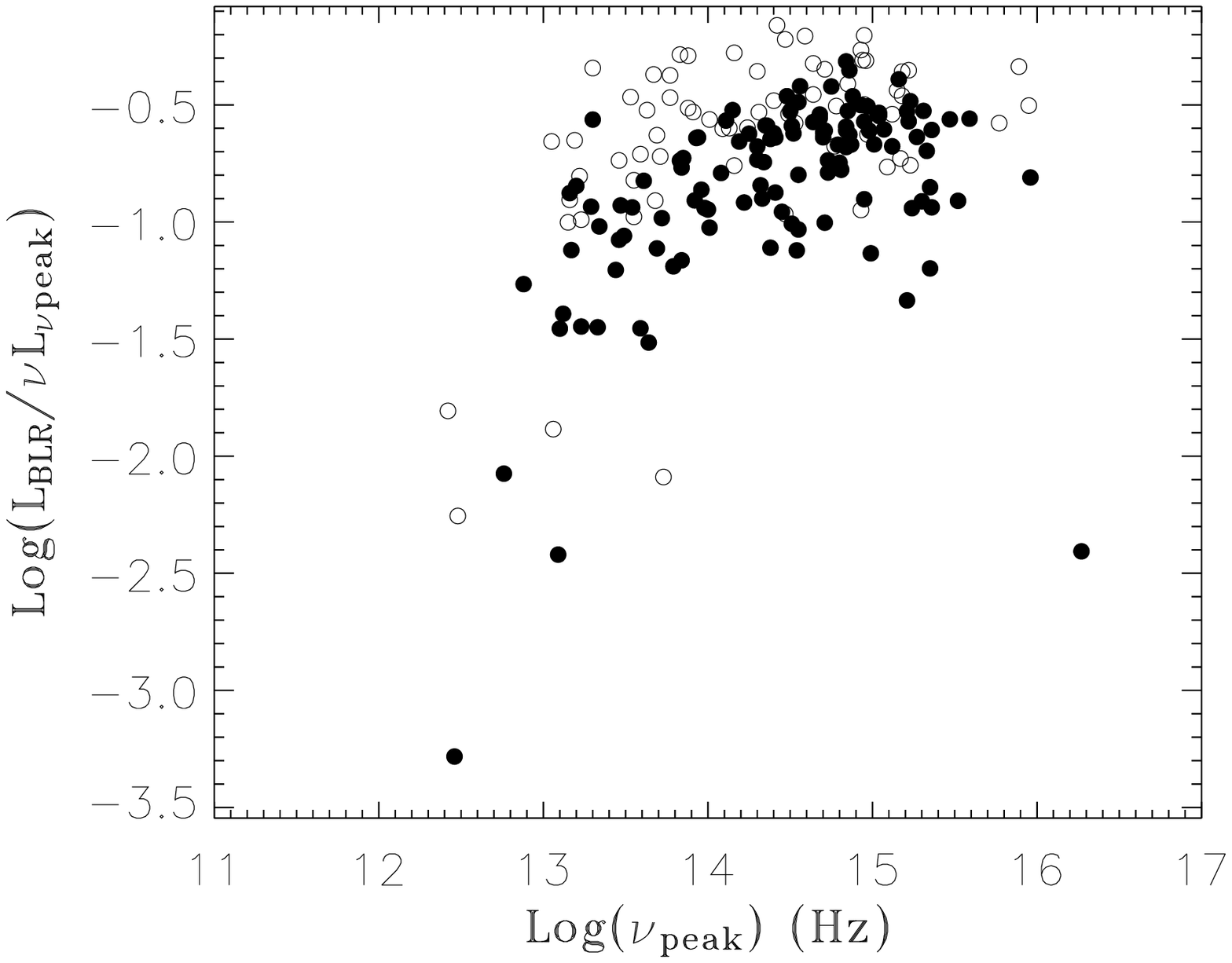}\\
\caption{The synchrotron peak frequency versus $\Luline/ \nuLupeak$.
The symbols are same as in Fig. \ref{rox}.} \label{nuLulinepeak.ps}
\end{figure}

We plot the ratio of BLR luminosity to synchrotron peak luminosity
$\Luline/ \nuLupeak$ and synchrotron peak frequency $\rm \nupeak$ in
Fig. \ref{nuLulinepeak.ps}, from which a significant correlation
between these two parameters is found with the Spearman correlation
coefficient $r=0.480$ at $\gg$ 99.99\% confidence level. However,
there is no strong correlation between the synchrotron peak
frequency and BLR luminosity.

\begin{figure}
\center
\includegraphics[width=8.5cm]{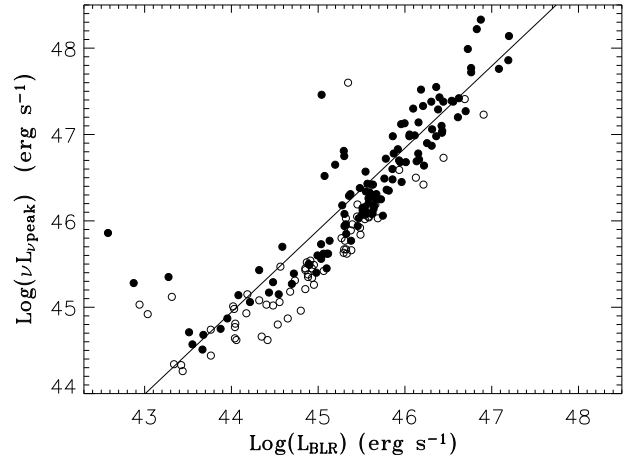}\\
\caption{The $\nuLupeak$ versus $L_{\rm BLR}$. The solid line is the
OLS bisector linear fit for 118 nonthermal-dominated FSRQs. The
symbols are same as in Fig. \ref{rox}.} \label{jd}
\end{figure}

The jet-disk relation can be further explored through the
relationship between $\nuLupeak$ and $L_{\rm BLR}$ shown in Fig.
\ref{jd}. A significant correlation is present with a Spearman
correlation coefficient of $r=0.909$ at confidence level
$\gg99.99\%$, which remains in partial correlation analysis to
exclude the common dependence on redshift. The ordinary least-square
(OLS) bisector linear fit gives,
\begin{equation}
\rm log~ \it \nuLupeak = \rm (0.95\pm0.06)~ \rm log~ \it L_{\rm BLR}
+ \rm (3.14\pm2.84)
\end{equation}
Our result implies a tight relation between jet and disk, which is
consistent with previous findings in various occasions, from the
strong correlations either between the radio emission and emission
line luminosity (e.g. Rawlings et al. 1989; Cao \& Jiang 2001), or
between the emission line luminosity and jet kinetic power in
different scales (Rawlings \& Saunders 1991; Celotti \& Fabian 1993;
Wang et al. 2004; Gu, Cao \& Jiang 2009). Moreover, our result is
consistent with the nearly proportional relation, $Q\propto L_{\rm
NLR}^{0.9\pm0.2}$, found between the jet bulk kinetic power and
narrow line luminosity (Rawlings \& Saunders 1991). However, it
should be noted that the synchrotron peak luminosity of blazars are
usually Doppler boosted, therefore, it may not be a good indicator
of jet power. It would be necessary to improve our result using
intrinsic parameters, either the intrinsic synchrotron peak
luminosity after eliminating the beaming effect, or
the jet power. %Nevertheless, our result will not be changed as long
%as the Doppler factor of our sample does not vary too much from
%source to source.

\section{discussion}

According to the blazar sequence, FSRQs with high synchrotron peak
frequency, e.g. $\nupeak>10^{15}$ Hz, and X-rays dominated by
synchrotron emission are not expected to exist. Due to its
importance, such objects have been extensively searched
\cite[see][and references therein]{padovani07}. The discoveries of
such blazars have been claimed, however, mainly on the basis of
their broad spectral indices, i.e. the ratios of radio/optical/X-ray
fluxes \citep{pad02,bas07,gio07}. In this sense, the X-ray
spectroscopy is required to confirm the nature of the X-ray emission
in these blazars. In its simplest form, a X-ray spectral index of
$\alpha<1$ indicates a inverse Compton origin, while $\alpha>1$ for
synchrotron X-ray emission. \cite{mar08} investigate the X-ray
spectra for a sample of 10 X-ray selected FSRQs from the Einstein
Medium Sensitivity Survey (EMSS) and four controversial sources
claimed to have synchrotron X-ray emission. They found that, in the
case of the EMSS broad line blazars, X-ray selection does not lead
to find sources with synchrotron peaks in the UV/X-ray range, as was
the case for X-ray-selected BL Lacs. Instead, for a wide range of
radio powers all the sources with broad emission lines show similar
SEDs, with synchrotron components peaking below the optical/UV
range. Moreover, the authors argued that four `anomalous' blazars
are no longer `anomalous' after a complete analysis of $Swift$ and
$INTEGRAL$ data, with two sources having inverse Compton X-rays, one
source being HBL, and the remaining one being narrow line Seyfert 1
galaxy without unambiguous evidence of X-ray emission from a
relativistic jet. Similarly, the $XMM-Newton$ and $Chandra$ X-ray
spectroscopy of 10 FSRQs were investigated by \cite{landt08}, which
are candidates to have an X-ray spectrum dominated by jet
synchrotron emission. However, the authors failed to find FSRQs with
X-ray spectra dominated by jet synchrotron emission, instead, the
X-rays are either from inverse Compton or are at transition between
the synchrotron and inverse Compton jet components as in IBLs. So,
despite the efforts to search for objects which may violate the
sequence trends no strong outliers have been found (Maraschi et al.
2008). As a severe challenge to blazar sequence, the selection of
important candidates of high $\nupeak$ luminous FSRQs are refined to
choose highly core-dominated radio quasars with low radio core to
X-ray luminosity ratios, e.g. log $(L_{\rm core}/L_{\rm X})\lesssim
5$ \citep{landt08}. The inverse Compton emission is expected to peak
at $\gamma$-ray frequencies, therefore, the high energy-peaked FSRQs
could be prime targets for the Fermi Gamma-ray Telescope
\citep{landt08}. Although the high energy-peaked FSRQs are not
firmly found yet, \cite{ghisel08} proposed a model to explain the
existence of blue quasars. In their scenario, the jet dissipation
region is out of the broad line region, resulting in a much reduced
energy density of BLR photons in jet region. Therefore, the cooling
due to the inverse Compton process is not severe, causing a high
synchrotron peak frequency though the source luminosity is high.
%ROXA sample is
%constructed by cross-correlating the SDSS, 2df and FIRST/GB6 (???),
%so our sources is contained by their sample ??? ROXA
%J081009.9+384757.0 have been shown to have inverse Compton X-rays
%\citep{mar08}, instead of synchrotron emission claimed by
%\cite{gio07}. And also 1629+4008 source is narrow line Seyfert 1
%galaxy \citep{mar08}. Though the $\gamma_{\rm peak}$ is high from
%the model fitting, it follows the trend of $\gamma_{\rm peak}$
%decreasing with increasing energy density of \cite{cel08} due to its
%low energy density. The VLBA images show a unresolved radio
%structure in 2.3, 5 and 8.4 GHz, and the spectrum between 2.3 and
%8.4 GHz is inverted (Gu \& Chen, in preparation). 1629+4008 has core
%dominance log $R>1.55$, and high brightness temperature is also
%found log $T_{\rm b}\gtrsim 11$, with especially log $T_{\rm
%b}=12.4$ and log $T_{\rm b,var}=12.2$ at 8.4 GHz (Gu \& Chen, in
%preparation). We therefore believe that it is reasonable to
%hypothesize that this object hosts a strong relativistic jet.
%However, whether jet dominated or NLS1 dominated in X-rays needs
%further explorations.

Although the anti-correlation between $\nupeak$ and $\nuLupeak$ for
the combined sample of 288 blazars (Fig. \ref{allnuLup.ps}) is
significant, it largely differs with the blazar sequence in its
significant scatter, which indicates that the synchrotron peak
luminosity can not explicitly determine the synchrotron peak
frequency, and vice verse. The advantages of using the synchrotron
peak luminosity is that the most of synchrotron emission are
radiated at synchrotron peak frequency, at which the luminosity can
be a good indicator of synchrotron emission. Since the synchrotron
peak frequency varies from source to source, the luminosity at fixed
wavebands (e.g. optical) is actually from the different portion of
source SED. The defect of the synchrotron peak luminosity lies in
the contamination from beaming effect, which precludes it to well
indicate the intrinsic source power. Only when the Doppler boosting
is known for each source, the intrinsic source power can be
obtained. This can not be performed at present stage. However, this
effect has been explored through eliminating the Doppler boosting in
blazar samples. Interestingly, recent studies have shown that the
negative correlation between $\nupeak$ and $\nuLupeak$ is likely an
artefact of Doppler boosting \citep{nieppola08,wu07}. According to
these authors, the negative correlation is not present when the
intrinsic parameters are used, conversely, a positive correlation is
claimed. The key point in these studies is the strong
anti-correlation between the Doppler factor and synchrotron peak
frequency in the way that the sources with their synchrotron peak at
low energies are significantly more boosted than high $\nupeak$
sources, which is found either from the variability Doppler factor
\citep{nieppola08} or from the one estimated with empirical relation
\citep{wu07}. %If the Doppler factor of our sample follow this trend,
%a positive correlation is highly possible.
%As outliers to blazar
%sequence, low - $\nupeak$ - low - power sources are clearly present
%\cite[see also][]{padovani07}, and the low - $\nupeak$ - low - power
%FSRQs are comparable to BL Lacs. In the Fig. \ref{allnuLup.ps} the
%high - $\nupeak$ FSRQs do not have high peak luminosity, and at same
%$\nupeak$ the peak luminosity of FSRQs, as a whole, are higher than
%that of BL Lacs, but are well covered by BL Lacs. We found $\nupeak
%> 10^{15}$ for 26 sources, nevertheless $\nupeak$ of these FSRQs do
%not reach the extreme $\nupeak$ of HBLs. The $\nupeak$ and
%$\nuLupeak$ of 170 BL Lac \citep{wu07} only have a weak negative
%correlation, however, it is significant for the combined sample of
%335 blazars.

The synchrotron peak frequency $\nupeak\propto B\delta\gamma_{\rm
peak}^2$, where $B$ is the magnetic field, $\delta$ the Doppler
factor, and $\gamma_{\rm peak}$ a characteristic electron energy
that is determined by a competition between accelerating and cooling
processes. Different from BL Lacs, the external inverse Compton
scattering is thought to be the dominant cooling process in FSRQs,
especially that upon BLR photons. %The anti-correlation between the
%$\gamma_{\rm peak}$ and the energy density were found and thus
%offered an explanation of the phenomenological blazar sequence
%\citep{ghisel98,ghi02}. \cite{cel08} recently estimate the jet power
%for a sample of blazars.
Through model fitting to blazar SEDs, the electron peak energy is
found to be well anti-correlated with the total energy density
(radiative and magnetic), which is thought to be the physics behind
the phenomenological blazar sequence \citep{ghisel98,ghi02,cel08}.
More energy density inside the jet cause a more severe cooling,
resulting in a smaller $\gamma_{\rm peak}$ then $\nupeak$. According
to the equations we used to estimate the black hole mass, we have
the BLR radius approximately with $R_{\rm BLR}\propto L_{\rm
BLR}^{0.6}$ for $\rm H\alpha, H\beta, Mg II, C IV$ broad lines.
Consequently, the energy density of BLR photons $u_{\rm BLR}^{*}$ is
expected to be proportional to $L_{\rm BLR}^{-0.2}$. We expect to
see an anti-correlation between $\nupeak$ and the BLR luminosity
$L_{\rm BLR}$. However, we failed to find any correlation between
$\nupeak$ and $L_{\rm BLR}$. Several factors may erase the expected
anti-correlation, e.g. the scatters in the derivation of BLR
luminosity from individual lines, the accuracy of empirical relation
in estimating BLR radius, and the inclusion of $B$ and $\delta$
(vary from source to source) in $\nupeak$. As a result, the positive
correlation between $\nupeak$ and $\Luline/\nuLupeak$ can be partly
(if not all) the result of the weak anti-correlation between
$\nupeak$ and $\nuLupeak$ (see Fig. \ref{nuLulinepeak.ps}).
Nevertheless, it indicates that FSRQs with higher ratio of disk to
jet emission could have higher peak
frequency. %\textbf{what's for BL Lacs ???}

%The strong correlation between $\Lubol/ \LuEdd$ and $\nupeak$
%implies a tight relation between the accretion process and the jet
%formation. The intrinsic nature of this correlation lies in the fact
%that the $\nupeak$ is determined from the SED, i.e. the overall
%multi-band emissions, while the Eddington ratio, the indicator of
%accretion rate, is tightly related with the properties of emission
%lines in SDSS spectra. The higher accretion rate results a higher
%jet power if the jet outflow rate is proportional to the accretion
%rate as stated in \cite{ghisel08}. A naive explanation of strong
%correlation could be that the electron distribution are forced to
%have higher $\gamma_{\rm peak}$ in order to possess higher jet
%power, assuming a similar power-law electron distribution in each
%FSRQ. There have been claims that the black hole of radio-loud AGNs
%are systematically heavier than that of radio quite ones
%\cite[e.g.][]{lao00} and the jet formation may be related with black
%hole mass. However, our results shown that the synchrotron peak
%frequency seems to not correlated with the black hole mass. Again,
%the synchrotron peak frequency is actually a combination of $B$,
%$\delta$ and $\gamma_{\rm peak}$, and the dependence of each
%parameters on the black hole mass is largely unknown, which
%precludes us to further explore the jet-black hole relation.

The FSRQs in our sample are defined from the flat spectrum between
1.4 and 4.85 GHz $\alpha<0.5$. However, this definition may
influenced by several factors. It is well known that FSRQs usually
show strong and rapid variability \cite[e.g.][]{gu06}. %Therefore,
%the non-simultaneous radio data may cause the uncertainty in source
%definition.
Another factor is the different resolution at 1.4 and
4.85 GHz. FIRST 1.4 GHz data are obtained from VLA observations,
which have much higher resolution than Green Bank telescope
observations for 4.85 GHz data. %For extended sources, more accurate
%flux measurements can be obtained from lower resolution radio
%observations, whereas a significant fraction of the flux can be
%missed from high resolution observations. This effect can be much
%less for the compact sources. Nevertheless, potentially some FSRQs
%of our sample could actually be SSRQs if using data with similar
%resolution.
In these respects, the simultaneous multi-band
observations with same telescope configuration (same resolution) is
required to calculate the radio spectral index, and then to
understand the nature of sources. %Generally, accretion disk emission

\section{Summary}

We have constructed a sample of 185 Flat Spectrum Radio Quasars
(FSRQs) by cross-correlating the \cite{Shen06} SDSS DR3 X-ray quasar
sample with FIRST and GB6 radio catalogues. From the spectrum energy
distraction (SED) constructed using multi-band (radio, optical,
Infrared and X-ray) data, we derived the synchrotron peak frequency
and peak luminosity. The black hole mass $M_{\rm BH}$ and the broad
line region (BLR) luminosity (then the bolometric luminosity $L_{\rm
bol}$) were obtained by measuring the line-width and strength of
broad emission lines from SDSS spectra. We define a subsample of 118
FSRQs, of which the nonthermal jet emission are thought to be
dominated over thermal ones from accretion disk and host galaxy. The
various correlations were explored for this subsample. The main
results are summarized below.

1. A weak anti-correlation is found between the synchrotron peak
frequency and peak luminosity. When combining our FSRQs sample with
the \cite{wu07} sample of 170 BL Lac objects, a significant
anti-correlation between the synchrotron peak frequency and
luminosity apparently presents covering about seven order of
magnitude in $\nupeak$. However, the anti-correlation differs with
the blazar sequence in the large scatter.

2. We found 25 FSRQs having synchrotron peak frequency $\nupeak >
10^{15}$ Hz, which is higher than the typical value for FSRQs. These
sources with high $\nupeak$ could be the targets for the Fermi
Gamma-ray telescope. At similar peak frequency, the peak luminosity
of FSRQs with $\nupeak > 10^{15}$ Hz is systematically higher than
that of BL Lac objects, with some FSRQs out of the range covered by
BL Lac objects. Though high $\nupeak$ are found in some FSRQs, they
do not reach the extreme value of BL Lacs.

3. No strong correlations are found either between the synchrotron
peak frequency and black hole mass, or between the synchrotron peak
frequency and the Eddington ratio. The peak luminosity is found to
be tightly correlated with both black hole mass and the Eddington
ratio indicating that the jet physics may be tightly related with
the accretion process, which is further confirmed by the tight
correlation between the synchrotron peak luminosity and the BLR
luminosity.

\section{acknowledgements}
We thank the referee, Hermine Landt, for insightful suggestions,
which is great helpful in improving our paper. We thank Tinggui
Wang, Xiaobo Dong, Dawei Xu, Wei Zhang, Tuo Ji and Chen Hu for helps
on data reduction. Z.Y. Chen thanks Center for Astrophysics of USTC
for hospitality during his stay in USTC. Shiyin Shen are appreciated
for providing the X-ray data, Marianne Vestergaard for Fe templates,
and Paolo Padovani for information on `HBL box'. We thank Markus
B\"{o}ttcher, Dongrong Jiang, Zhonghui Fan and Fuguo Xie for useful
discussions. This work makes use of data products from the Two
Micron All Sky Survey, which is a joint project of the University of
Massachusetts and the Infrared Processing and Analysis
Center/California Institute of Technology, funded by the National
Aeronautics and Space Administration and the National Science
Foundation. This work is supported by National Science Foundation of
China (grants 10633010, 10703009, 10833002, 10773020 and 10821302),
973 Program (No. 2009CB824800), and the CAS (KJCX2-YW-T03).

Funding for the SDSS and SDSS-II has been provided by the Alfred P.
Sloan Foundation, the Participating Institutions, the National
Science Foundation, the U.S. Department of Energy, the National
Aeronautics and Space Administration, the Japanese Monbukagakusho,
the Max Planck Society, and the Higher Education Funding Council for
England. The SDSS Web Site is http://www.sdss.org/.

The SDSS is managed by the Astrophysical Research Consortium for the
Participating Institutions. The Participating Institutions are the
American Museum of Natural History, Astrophysical Institute Potsdam,
University of Basel, University of Cambridge, Case Western Reserve
University, University of Chicago, Drexel University, Fermilab, the
Institute for Advanced Study, the Japan Participation Group, Johns
Hopkins University, the Joint Institute for Nuclear Astrophysics,
the Kavli Institute for Particle Astrophysics and Cosmology, the
Korean Scientist Group, the Chinese Academy of Sciences (LAMOST),
Los Alamos National Laboratory, the Max-Planck-Institute for
Astronomy (MPIA), the Max-Planck-Institute for Astrophysics (MPA),
New Mexico State University, Ohio State University, University of
Pittsburgh, University of Portsmouth, Princeton University, the
United States Naval Observatory, and the University of Washington.

%%%%%%%%%%%%%%%%%%%%%%%%%%%%%%%%%%%%%%%%%%%%%%%%%%%%%%%%%%%%%%%%%%%%%%%%%%%%%%%%%%%%%%%%%%
\begin{table*}
\setlength{\tabcolsep}{0.08in}
 %\tabletypesize{\scriptsize}
 \caption{Nonthermal jet-dominated FSRQs. (1) Source SDSS name; (2) redshift;
 (3) black hole mass; (4) synchrotron peak luminosity; (5)broad line region luminosity;
 (6) bolometric luminosity; (7) synchrotron peak frequency;
 spectral indices between the rest-frame frequencies of (8) 5 GHz
 and 5000 $\rm \AA$, (9) 5000 $\rm \AA$ and 1 keV, (10) 5 GHz and 1
 keV. \label{tbl-1}}
 \begin{center}
\begin{tabular}{lccccccccccc}
\hline\hline SDSS name &  $z$ & ${\rm Log}~M_{\rm BH}$ & ${\rm
Log}~\nu L_{\nu {\rm peak}}$  &  ${\rm Log}~ L_{\rm BLR}$    &
 ${\rm Log}~ L_{\rm bol}$    &  ${\rm Log}~ \nu_{\rm peak}$   &  $\alpha_{\rm ro}$
  & $\alpha_{\rm ox}$   &  $\alpha_{\rm rx}$   \\
% \hline
    &     & ($M_\odot$)   & ($\rm erg\,s^{-1}$)   & ($\rm erg\,s^{-1}$)  & ($\rm erg\,s^{-1}$) &  (Hz)        \\
%\hline
(1) & (2) & (3)  & (4)  & (5)  &  (6) & (7) & (8) & (9)  & (10) \\
\hline
J001130.4+005752.1     &  1.492  &   8.36  &  46.18  &  45.28  &  46.28  &  14.95  &  0.555  &  1.306  &  0.809   \\
J013352.7+011343.6     &  0.308  &   8.56  &  44.75  &  43.88  &  44.88  &  14.41  &  0.475  &  1.288  &  0.750   \\
J074541.7+314256.6     &  0.461  &   9.66  &  46.14  &  45.64  &  46.64  &  14.93  &  0.521  &  1.666  &  0.909   \\
J074625.9+254902.2     &  2.979  &   9.69  &  47.42  &  46.62  &  47.62  &  14.55  &  0.574  &  1.389  &  0.850   \\
J081009.9+384756.9     &  3.946  &   9.33  &  47.76  &  47.08  &  48.08  &  15.12  &  0.050  &  1.882  &  0.671   \\
J081303.8+254211.1     &  2.024  &   9.68  &  47.30  &  46.10  &  47.10  &  13.44  &  0.508  &  1.377  &  0.802   \\
J082455.5+391641.8     &  1.216  &   8.71  &  46.81  &  45.30  &  46.30  &  13.64  &  0.599  &  1.539  &  0.918   \\
J083148.9+042939.2     &  0.174  &   7.42  &  45.86  &  42.58  &  43.58  &  12.46  &  0.602  &  1.303  &  0.840   \\
J083636.9+412554.9     &  1.301  &   9.29  &  46.78  &  45.87  &  46.87  &  15.52  &  0.473  &  1.183  &  0.714   \\
J085220.5+473458.4     &  2.420  &   9.25  &  46.83  &  45.92  &  46.92  &  15.30  &  0.517  &  1.108  &  0.718   \\
J085442.0+575730.1     &  1.317  &   9.59  &  46.99  &  46.11  &  47.11  &  13.16  &  0.590  &  1.475  &  0.890   \\
J090304.0+465104.5     &  1.470  &   9.47  &  46.68  &  45.95  &  46.95  &  14.30  &  0.622  &  1.497  &  0.919   \\
J090910.1+012136.0     &  1.024  &   9.33  &  46.72  &  45.78  &  46.78  &  13.98  &  0.515  &  1.517  &  0.855   \\
J092058.5+444153.9     &  2.190  &   9.84  &  47.99  &  46.72  &  47.72  &  12.88  &  0.446  &  1.502  &  0.804   \\
J092656.7+041613.0     &  1.036  &  10.14  &  46.45  &  45.96  &  46.96  &  14.55  &  0.300  &  1.242  &  0.619   \\
J093035.1+464408.6     &  2.033  &   9.38  &  47.13  &  46.00  &  47.00  &  14.99  &  0.420  &  1.540  &  0.800   \\
J093200.1+553347.0     &  0.266  &   8.92  &  44.87  &  43.95  &  44.95  &  14.22  &  0.420  &  1.514  &  0.791   \\
J093256.8+074211.9     &  1.004  &   9.23  &  46.15  &  45.51  &  46.51  &  14.41  &  0.285  &  1.479  &  0.690   \\
J093309.3+461534.6     &  0.778  &   9.39  &  45.73  &  45.03  &  46.03  &  15.33  &  0.414  &  1.303  &  0.715   \\
J094855.4+403944.7     &  1.249  &   9.39  &  46.60  &  45.86  &  46.86  &  14.34  &  0.608  &  1.513  &  0.915   \\
J094857.3+002225.6     &  0.585  &   8.41  &  45.43  &  44.32  &  45.32  &  14.38  &  0.609  &  1.343  &  0.858   \\
J095556.4+061642.5     &  1.278  &   9.54  &  46.98  &  45.86  &  46.86  &  13.17  &  0.334  &  1.428  &  0.705   \\
J095738.2+552257.9     &  0.895  &   9.11  &  46.52  &  45.07  &  46.07  &  13.23  &  0.669  &  1.532  &  0.961   \\
J095819.7+472507.7     &  1.882  &   9.52  &  47.14  &  46.16  &  47.16  &  13.72  &  0.590  &  1.451  &  0.882   \\
J095855.1+423703.8     &  0.664  &   8.80  &  45.39  &  44.72  &  45.72  &  14.79  &  0.553  &  1.189  &  0.769   \\
J100556.0+433238.5     &  1.587  &   8.94  &  46.29  &  45.36  &  46.36  &  13.29  &  0.521  &  1.366  &  0.808   \\
J101027.5+413238.9     &  0.612  &   9.56  &  46.10  &  45.56  &  46.56  &  14.68  &  0.570  &  1.376  &  0.843   \\
J101122.6+470042.2     &  2.928  &   9.46  &  47.52  &  46.19  &  47.19  &  15.21  &  0.249  &  1.506  &  0.675   \\
J101557.1+010913.7     &  0.780  &   9.85  &  46.36  &  45.79  &  46.79  &  14.95  &  0.449  &  1.740  &  0.887   \\
J101725.9+611627.5     &  2.805  &  10.17  &  47.77  &  46.76  &  47.76  &  14.51  &  0.468  &  1.558  &  0.837   \\
J102235.6+454105.4     &  0.743  &   8.63  &  45.06  &  44.21  &  45.21  &  13.20  &  0.564  &  1.318  &  0.819   \\
J102429.0+052952.4     &  1.482  &   9.65  &  46.98  &  46.05  &  47.05  &  13.47  &  0.251  &  1.694  &  0.740   \\
J102713.1+480313.5     &  1.286  &   9.23  &  46.31  &  45.37  &  46.37  &  13.54  &  0.570  &  1.385  &  0.846   \\
J103025.0+551622.3     &  0.434  &   8.81  &  45.50  &  44.89  &  45.89  &  14.98  &  0.387  &  1.560  &  0.785   \\
J103050.9+531028.5     &  1.197  &   9.41  &  46.48  &  45.86  &  46.86  &  14.25  &  0.216  &  1.487  &  0.647   \\
J103144.7+602030.7     &  1.230  &   9.21  &  46.43  &  45.57  &  46.57  &  13.96  &  0.495  &  1.462  &  0.823   \\
J103351.4+605107.5     &  1.401  &   9.15  &  46.65  &  45.20  &  46.20  &  13.59  &  0.546  &  1.573  &  0.894   \\
J104146.8+523328.4     &  0.678  &   9.53  &  46.07  &  45.54  &  46.54  &  15.21  &  0.576  &  1.492  &  0.886   \\
J104148.9+523355.6     &  2.301  &   9.49  &  48.33  &  46.87  &  47.87  &  13.10  &  0.336  &  1.688  &  0.794   \\
J104410.6+532220.6     &  1.901  &   9.44  &  47.12  &  45.96  &  46.96  &  13.84  &  0.477  &  1.512  &  0.828   \\
J104542.2+525112.6     &  1.058  &   9.33  &  46.30  &  45.64  &  46.64  &  14.19  &  0.409  &  1.496  &  0.777   \\
J104552.7+062436.3     &  1.509  &  10.02  &  47.38  &  46.30  &  47.30  &  13.46  &  0.424  &  1.719  &  0.863   \\
J104732.3+483531.3     &  0.866  &   8.93  &  45.60  &  44.99  &  45.99  &  14.84  &  0.508  &  1.361  &  0.797   \\
J105654.2+051713.2     &  0.456  &   8.94  &  45.15  &  44.55  &  45.55  &  15.07  &  0.484  &  1.326  &  0.770   \\
J111914.3+600457.2     &  2.646  &  10.43  &  48.14  &  47.20  &  48.20  &  15.24  &  0.275  &  1.459  &  0.677   \\
J112115.7+575446.8     &  1.274  &   9.24  &  46.34  &  45.55  &  46.55  &  14.08  &  0.293  &  1.383  &  0.663   \\
J113245.6+003427.7     &  1.223  &   9.09  &  47.46  &  45.04  &  46.04  &  13.09  &  0.384  &  1.836  &  0.876   \\
J114521.3+045526.7     &  1.343  &   9.37  &  46.20  &  45.58  &  46.58  &  14.40  &  0.596  &  1.428  &  0.878   \\
J120127.4+090040.4     &  1.016  &   9.90  &  46.69  &  46.13  &  47.13  &  14.68  &  0.233  &  1.365  &  0.617   \\
J121201.6+605034.6     &  1.146  &   9.38  &  46.25  &  45.66  &  46.66  &  14.35  &  0.314  &  1.490  &  0.713   \\
J121347.5+000130.1     &  0.962  &   9.40  &  46.18  &  45.65  &  46.65  &  15.04  &  0.434  &  1.373  &  0.752   \\
J122106.9+454852.2     &  0.525  &   9.26  &  45.62  &  45.06  &  46.06  &  15.59  &  0.429  &  1.234  &  0.702   \\
J122339.3+461118.7     &  1.013  &   9.26  &  46.33  &  45.59  &  46.59  &  13.83  &  0.521  &  1.367  &  0.808   \\
J122452.4+033050.3     &  0.956  &   9.24  &  46.03  &  45.47  &  46.47  &  13.30  &  0.770  &  1.374  &  0.975   \\
J123530.6+522828.0     &  1.651  &   9.70  &  47.00  &  46.05  &  47.05  &  14.00  &  0.317  &  1.590  &  0.749   \\
J123932.8+044305.4     &  1.762  &   9.18  &  46.31  &  45.67  &  46.67  &  13.94  &  0.627  &  1.165  &  0.809   \\
J125500.5+034043.2     &  0.437  &   8.86  &  45.29  &  44.48  &  45.48  &  15.96  &  0.524  &  0.998  &  0.685   \\
J125545.1+612450.8     &  2.056  &   9.56  &  46.70  &  45.93  &  46.93  &  13.84  &  0.439  &  1.251  &  0.714   \\
J130217.2+481917.5     &  0.874  &   9.09  &  46.07  &  45.50  &  46.50  &  14.11  &  0.556  &  1.320  &  0.815   \\
J130603.4+552943.7     &  1.601  &   9.53  &  46.90  &  46.25  &  47.25  &  14.38  &  0.473  &  1.489  &  0.818   \\

\hline
\end{tabular}
\end{center}
\end{table*}
\addtocounter{table}{-1}
%%%%%%%%%%%%%%%%%%%%%%%%%%%%%%%%%%%%%%%%%%%%%%%%%%%%%%%%%%%%%%%%%%%%%%%%%%%%%%%%%%
\begin{table*}
\setlength{\tabcolsep}{0.08in} \caption{Continued\dots
\label{tablesample}}
 \begin{center}
\begin{tabular}{lccccccccccc}
\hline\hline SDSS name &  $z$ & ${\rm Log}~M_{\rm BH}$ & ${\rm
Log}~\nu L_{\nu {\rm peak}}$  &  ${\rm Log}~ L_{\rm BLR}$    &
 ${\rm Log}~ L_{\rm bol}$  &  ${\rm Log}~ \nu_{\rm peak}$    &  $\alpha_{\rm ro}$
  & $\alpha_{\rm ox}$   &  $\alpha_{\rm rx}$   \\
% \hline
    &     & ($M_\odot$)   & ($\rm erg\,s^{-1}$)   & ($\rm erg\,s^{-1}$)  & ($\rm erg\,s^{-1}$) &  (Hz)        \\
%\hline
(1) & (2) & (3)  & (4)  & (5)  &  (6) & (7) & (8) & (9)  & (10)  \\
\hline

J130909.8+555738.5     &  1.629  &   9.64  &  47.06  &  46.31  &  47.31  &  14.80  &  0.450  &  1.665  &  0.861   \\
J133245.3+472222.6     &  0.669  &   8.62  &  45.70  &  44.59  &  45.59  &  13.69  &  0.591  &  1.441  &  0.879   \\
J133437.4+563147.9     &  0.343  &   8.11  &  44.51  &  43.67  &  44.67  &  14.32  &  0.584  &  1.403  &  0.862   \\
J133749.6+550102.3     &  1.099  &   9.00  &  46.38  &  45.48  &  46.48  &  14.33  &  0.565  &  1.640  &  0.930   \\
J133802.8+423957.2     &  2.237  &   9.56  &  47.27  &  46.70  &  47.70  &  15.22  &  0.241  &  1.393  &  0.632   \\
J134357.6+575442.6     &  0.933  &   8.93  &  45.77  &  45.13  &  46.13  &  14.70  &  0.598  &  1.346  &  0.852   \\
J134934.7+534116.8     &  0.979  &   9.65  &  46.35  &  45.81  &  46.81  &  15.03  &  0.617  &  1.437  &  0.895   \\
J135054.6+052206.6     &  0.442  &   9.30  &  45.27  &  44.70  &  45.70  &  14.64  &  0.457  &  1.470  &  0.800   \\
J135351.6+015153.3     &  1.608  &   9.45  &  46.64  &  46.22  &  47.22  &  14.56  &  0.509  &  1.454  &  0.829   \\
J135726.4+001543.8     &  0.662  &   9.11  &  45.45  &  45.10  &  46.10  &  14.86  &  0.579  &  1.271  &  0.813   \\
J135817.6+575204.4     &  1.373  &   9.72  &  47.29  &  46.38  &  47.38  &  13.92  &  0.210  &  1.693  &  0.713   \\
J140126.2+520834.6     &  2.972  &   9.69  &  47.86  &  47.19  &  48.19  &  15.01  &  0.219  &  1.626  &  0.696   \\
J141159.7+423950.4     &  0.886  &   9.60  &  46.25  &  45.72  &  46.72  &  14.85  &  0.382  &  1.629  &  0.805   \\
J141324.2+530525.7     &  0.456  &   8.10  &  44.57  &  43.55  &  44.55  &  13.34  &  0.626  &  1.289  &  0.851   \\
J141708.2+460705.6     &  1.559  &   9.68  &  47.02  &  46.43  &  47.43  &  14.84  &  0.534  &  1.655  &  0.914   \\
J142020.7+462441.0     &  1.245  &   9.50  &  45.94  &  45.30  &  46.30  &  15.27  &  0.385  &  1.370  &  0.719   \\
J145247.4+473529.1     &  1.158  &   9.48  &  46.08  &  45.62  &  46.62  &  14.88  &  0.279  &  1.288  &  0.621   \\
J145859.4+041614.1     &  0.392  &   8.20  &  45.35  &  43.28  &  44.28  &  12.76  &  0.669  &  1.502  &  0.951   \\
J150031.8+483646.8     &  1.028  &   9.86  &  46.87  &  46.31  &  47.31  &  15.47  &  0.121  &  1.188  &  0.483   \\
J150324.7+475830.4     &  0.345  &   7.49  &  45.28  &  42.87  &  43.87  &  16.27  &  0.400  &  0.965  &  0.591   \\
J150759.7+041512.1     &  1.701  &   9.61  &  47.33  &  46.21  &  47.21  &  14.54  &  0.303  &  1.848  &  0.827   \\
J152544.7+354446.7     &  1.101  &   9.16  &  45.96  &  45.32  &  46.32  &  13.93  &  0.356  &  1.394  &  0.708   \\
J153107.4+584410.0     &  1.724  &  10.00  &  47.20  &  46.61  &  47.61  &  14.36  &  0.265  &  1.759  &  0.771   \\
J153232.3+513001.5     &  1.876  &   9.68  &  47.10  &  46.42  &  47.42  &  14.30  &  0.329  &  1.726  &  0.803   \\
J153404.8+482340.7     &  0.543  &   8.52  &  45.17  &  44.43  &  45.43  &  14.73  &  0.605  &  1.408  &  0.877   \\
J153432.6+492049.2     &  1.294  &   9.41  &  46.49  &  45.76  &  46.76  &  13.85  &  0.414  &  1.561  &  0.803   \\
J153457.2+583923.5     &  1.901  &   9.28  &  47.04  &  46.43  &  47.43  &  14.71  &  0.368  &  1.301  &  0.684   \\
J154502.8+513500.9     &  1.930  &  10.05  &  47.39  &  46.54  &  47.54  &  15.35  &  0.483  &  1.447  &  0.810   \\
J154917.4+503805.9     &  2.175  &   9.84  &  47.38  &  46.56  &  47.56  &  13.61  &  0.561  &  1.277  &  0.804   \\
J154929.4+023701.1     &  0.414  &   8.95  &  45.40  &  44.98  &  45.98  &  14.75  &  0.667  &  1.576  &  0.975   \\
J160226.9+274141.6     &  0.938  &   9.45  &  46.09  &  45.63  &  46.63  &  14.48  &  0.481  &  1.425  &  0.801   \\
J160623.6+540555.7     &  0.876  &   9.40  &  46.12  &  45.51  &  46.51  &  15.36  &  0.483  &  1.264  &  0.748   \\
J160658.3+271705.9     &  0.934  &   9.01  &  45.77  &  45.38  &  46.38  &  15.16  &  0.627  &  1.311  &  0.859   \\
J160822.1+401217.9     &  0.628  &   8.33  &  45.14  &  44.08  &  45.08  &  13.49  &  0.655  &  1.322  &  0.881   \\
J160913.2+535429.7     &  0.992  &   9.72  &  46.16  &  45.54  &  46.54  &  14.52  &  0.471  &  1.346  &  0.768   \\
J161447.0+374607.3     &  1.532  &   9.77  &  47.43  &  46.40  &  47.40  &  14.55  &  0.183  &  1.992  &  0.796   \\
J162229.3+400643.5     &  0.688  &   9.04  &  45.49  &  44.90  &  45.90  &  14.51  &  0.519  &  1.267  &  0.773   \\
J162307.6+390932.5     &  1.975  &   9.90  &  47.72  &  46.76  &  47.76  &  14.45  &  0.336  &  1.824  &  0.841   \\
J162330.6+355933.1     &  0.866  &   9.32  &  45.85  &  45.32  &  46.32  &  14.50  &  0.574  &  1.478  &  0.881   \\
J162422.0+392441.0     &  1.116  &   9.13  &  46.26  &  45.58  &  46.58  &  14.84  &  0.438  &  1.348  &  0.746   \\
J162553.3+434713.8     &  1.048  &   8.79  &  46.08  &  45.30  &  46.30  &  14.81  &  0.466  &  1.227  &  0.724   \\
J162901.3+400759.6     &  0.272  &   8.06  &  44.71  &  43.51  &  44.51  &  15.35  &  0.363  &  1.405  &  0.716   \\
J163302.1+392427.6     &  1.024  &   9.51  &  46.68  &  46.01  &  47.01  &  14.87  &  0.258  &  1.757  &  0.766   \\
J163515.5+380804.6     &  1.814  &   9.83  &  47.55  &  46.36  &  47.36  &  13.79  &  0.582  &  1.861  &  1.016   \\
J163709.3+414030.6     &  0.760  &  10.05  &  46.06  &  45.75  &  46.75  &  14.84  &  0.299  &  1.527  &  0.715   \\
J163755.2+250930.5     &  1.107  &   9.50  &  46.42  &  45.63  &  46.63  &  14.73  &  0.295  &  1.460  &  0.690   \\
J164258.8+394837.2     &  0.593  &   9.21  &  46.75  &  45.30  &  46.30  &  13.33  &  0.656  &  1.897  &  1.077   \\
J164829.2+410405.9     &  0.852  &   9.38  &  45.62  &  45.10  &  46.10  &  14.15  &  0.618  &  1.222  &  0.823   \\
J165005.5+414032.6     &  0.585  &   9.05  &  45.56  &  45.03  &  46.03  &  15.31  &  0.529  &  1.296  &  0.789   \\
J165201.5+623209.3     &  1.633  &   9.00  &  46.57  &  45.55  &  46.55  &  14.01  &  0.394  &  1.468  &  0.758   \\
J165329.9+310756.9     &  1.298  &   9.55  &  46.71  &  46.16  &  47.16  &  15.04  &  0.406  &  1.588  &  0.806   \\
J165931.9+373529.0     &  0.771  &   9.59  &  45.94  &  45.46  &  46.46  &  15.23  &  0.371  &  1.467  &  0.742   \\
J170232.6+315752.2     &  1.951  &   9.89  &  47.38  &  46.44  &  47.44  &  15.36  &  0.229  &  1.542  &  0.674   \\
J170425.1+333146.1     &  0.290  &   8.30  &  44.68  &  43.68  &  44.68  &  14.71  &  0.372  &  1.431  &  0.731   \\
J170648.1+321422.7     &  1.070  &   9.69  &  46.78  &  46.15  &  47.15  &  14.86  &  0.231  &  1.478  &  0.654   \\
J171715.2+262149.1     &  1.934  &   9.88  &  46.98  &  46.36  &  47.36  &  14.70  &  0.357  &  1.510  &  0.748   \\
J172732.4+584634.1     &  0.844  &   9.03  &  45.62  &  45.11  &  46.11  &  14.97  &  0.579  &  1.343  &  0.838   \\
J213638.6+004154.5     &  1.941  &   9.92  &  48.22  &  46.83  &  47.83  &  13.12  &  0.610  &  1.728  &  0.989   \\

\hline
\end{tabular}
\end{center}
\end{table*}
%\addtocounter{table}{-1}
%%%%%%%%%%%%%%%%%%%%%%%%%%%%%%%%%%%%%%%%%%%%%%%%%%%%%%%%%%%%%%%%%%%%%%%%%%%%%%%

%%{  Notes: Col. 1: the source name. Col. 2: the redshift. Col. 3: the black hole mass
%%in units of solar mass. Col. 4: the intrinsic synchrotron peak luminosity.  Col. 5:
%%the bolmatric luminosity.  Col. 6: the the intrinsic synchrotron peak frequency,
%%in units of Hz. The Col. 4, 5,  in units of $\rm erg~ s^{-1}$. }

\clearpage

\begin{table*}
\setlength{\tabcolsep}{0.08in}
 %\tabletypesize{\scriptsize}
 \caption{Thermal-dominated FSRQs. (1) Source SDSS name; (2) redshift;
 (3) black hole mass; (4) synchrotron peak luminosity; (5)broad line region luminosity;
 (6) bolometric luminosity; (7) synchrotron peak frequency;
 spectral indices between the rest-frame frequencies of (8) 5 GHz
 and 5000 $\rm \AA$, (9) 5000 $\rm \AA$ and 1 keV, (10) 5 GHz and 1
 keV. \label{tb2-2}}
 \begin{center}
\begin{tabular}{lccccccccccc}
\hline\hline SDSS name &  $z$ & ${\rm Log}~M_{\rm BH}$ & ${\rm
Log}~\nu L_{\nu {\rm peak}}$  &  ${\rm Log}~ L_{\rm BLR}$    &
 ${\rm Log}~ L_{\rm bol}$  &  ${\rm Log}~ \nu_{\rm peak}$   &  $\alpha_{\rm ro}$
  & $\alpha_{\rm ox}$   &  $\alpha_{\rm rx}$   \\
% \hline
    &     & ($M_\odot$)   & ($\rm erg\,s^{-1}$)   & ($\rm erg\,s^{-1}$)  & ($\rm erg\,s^{-1}$) &  (Hz)        \\
%\hline
(1) & (2) & (3)  & (4)  & (5)  &  (6) & (7) & (8) & (9)   & (10)  \\
\hline

J021225.6+010056.1     &  0.513  &   9.19  &  45.31  &  44.73  &  45.73  &  15.77  &  0.468  &  1.111  &  0.686   \\
J074218.2+194719.7     &  0.657  &   9.71  &  45.84  &  45.49  &  46.49  &  15.22  &  0.395  &  1.498  &  0.769   \\
J074242.2+374402.1     &  0.806  &   9.51  &  45.67  &  45.30  &  46.30  &  13.67  &  0.467  &  1.117  &  0.688   \\
J074559.3+331334.5     &  0.610  &   9.26  &  45.42  &  45.06  &  46.06  &  14.30  &  0.569  &  1.401  &  0.851   \\
J075448.8+303355.1     &  0.795  &   9.74  &  46.04  &  45.58  &  46.58  &  14.64  &  0.467  &  1.239  &  0.728   \\
J080131.9+473616.1     &  0.157  &   8.96  &  44.93  &  44.17  &  45.17  &  15.23  &  0.353  &  1.541  &  0.756   \\
J080644.4+484149.1     &  0.370  &   9.31  &  45.15  &  44.18  &  45.18  &  14.47  &  0.584  &  1.443  &  0.876   \\
J080856.6+405244.9     &  1.418  &   9.38  &  46.59  &  45.93  &  46.93  &  13.05  &  0.622  &  1.313  &  0.856   \\
J081059.0+413402.6     &  0.507  &   8.74  &  45.47  &  44.56  &  45.56  &  13.16  &  0.614  &  1.420  &  0.887   \\
J081409.2+323731.9     &  0.844  &   9.37  &  45.67  &  45.32  &  46.32  &  14.71  &  0.618  &  1.120  &  0.788   \\
J081916.5+264203.2     &  0.526  &   8.89  &  45.02  &  44.48  &  45.48  &  15.12  &  0.582  &  1.277  &  0.817   \\
J083353.9+422401.8     &  0.249  &   7.05  &  45.03  &  42.94  &  43.94  &  13.73  &  0.568  &  1.501  &  0.884   \\
J084203.7+401831.4     &  0.152  &   8.66  &  44.44  &  43.76  &  44.76  &  14.77  &  0.379  &  1.532  &  0.770   \\
J084650.0+064148.9     &  0.616  &   9.69  &  45.77  &  45.31  &  46.31  &  15.18  &  0.463  &  1.360  &  0.767   \\
J085040.0+543753.2     &  0.367  &   9.58  &  45.01  &  44.02  &  45.02  &  13.23  &  0.495  &  1.434  &  0.813   \\
J090835.8+415046.0     &  0.734  &   9.05  &  45.49  &  44.95  &  45.95  &  14.49  &  0.637  &  1.284  &  0.856   \\
J091401.8+050750.6     &  0.301  &   9.81  &  45.08  &  44.32  &  45.32  &  14.16  &  0.535  &  1.572  &  0.886   \\
J092554.7+400414.2     &  0.471  &   9.53  &  45.38  &  44.88  &  45.88  &  14.95  &  0.356  &  1.421  &  0.717   \\
J092703.0+390220.7     &  0.695  &   9.68  &  46.42  &  45.62  &  46.62  &  13.22  &  0.775  &  1.531  &  1.031   \\
J093712.3+500852.0     &  0.276  &   9.18  &  44.92  &  43.04  &  44.04  &  13.06  &  0.604  &  1.229  &  0.816   \\
J100017.6+000523.9     &  0.905  &   9.31  &  46.03  &  45.69  &  46.69  &  13.30  &  0.717  &  1.186  &  0.876   \\
J102106.0+452331.6     &  0.364  &   9.15  &  44.77  &  44.04  &  45.04  &  15.17  &  0.533  &  1.307  &  0.795   \\
J102738.5+605016.7     &  0.332  &   9.80  &  44.98  &  44.03  &  45.03  &  14.93  &  0.356  &  1.254  &  0.660   \\
J103214.6+635950.2     &  0.556  &   9.18  &  45.06  &  44.55  &  45.55  &  14.78  &  0.448  &  1.362  &  0.758   \\
J104207.6+501322.1     &  1.265  &   9.46  &  47.60  &  45.34  &  46.34  &  12.48  &  0.288  &  1.534  &  0.710   \\
J110539.0+020257.2     &  0.105  &   8.01  &  44.26  &  43.44  &  44.44  &  13.55  &  0.539  &  1.551  &  0.882   \\
J110718.9+100418.1     &  0.633  &   9.38  &  45.35  &  44.88  &  45.88  &  13.53  &  0.564  &  1.250  &  0.797   \\
J112023.2+540426.9     &  0.923  &   9.09  &  45.63  &  45.29  &  46.29  &  15.89  &  0.666  &  0.907  &  0.748   \\
J113251.1+541031.6     &  1.624  &   9.66  &  46.42  &  46.21  &  47.21  &  14.59  &  0.498  &  1.196  &  0.735   \\
J114803.2+565411.4     &  0.451  &   9.14  &  45.21  &  44.85  &  45.85  &  15.18  &  0.463  &  1.388  &  0.776   \\
J115542.5+021411.3     &  0.873  &   9.59  &  46.02  &  45.54  &  46.54  &  14.40  &  0.520  &  1.494  &  0.850   \\
J120822.5+524013.5     &  0.432  &   9.25  &  45.34  &  44.93  &  45.93  &  14.85  &  0.370  &  1.551  &  0.770   \\
J122940.9+624255.0     &  0.967  &   9.56  &  46.05  &  45.45  &  46.45  &  14.09  &  0.414  &  1.430  &  0.759   \\
J123157.1+542028.8     &  0.516  &   9.27  &  44.80  &  44.53  &  45.53  &  14.93  &  0.506  &  1.228  &  0.751   \\
J123807.8+532555.8     &  0.347  &   9.88  &  45.03  &  44.40  &  45.40  &  14.97  &  0.449  &  1.479  &  0.799   \\
J124116.5+514130.1     &  0.823  &   9.79  &  46.05  &  45.58  &  46.58  &  13.77  &  0.490  &  1.278  &  0.757   \\
J124139.7+493405.6     &  0.474  &   9.36  &  45.43  &  44.85  &  45.85  &  14.53  &  0.404  &  1.352  &  0.725   \\
J130444.0+013230.9     &  2.288  &   9.06  &  46.50  &  46.13  &  47.13  &  13.77  &  0.422  &  1.229  &  0.696   \\
J130554.2+014929.9     &  0.733  &   9.91  &  45.96  &  45.39  &  46.39  &  14.66  &  0.490  &  1.572  &  0.856   \\
J131211.1+480925.2     &  0.715  &   9.86  &  45.92  &  45.48  &  46.48  &  15.15  &  0.462  &  1.484  &  0.808   \\
J131827.1+620036.2     &  0.307  &   8.37  &  44.74  &  43.76  &  44.76  &  13.55  &  0.545  &  1.369  &  0.824   \\
J132631.5+473755.8     &  0.682  &   9.26  &  45.45  &  44.93  &  45.93  &  13.63  &  0.516  &  1.305  &  0.784   \\
J133253.3+020045.4     &  0.216  &   8.14  &  45.12  &  43.31  &  44.31  &  12.42  &  0.760  &  1.335  &  0.955   \\
J134617.6+622044.7     &  0.117  &   8.16  &  44.34  &  43.34  &  44.34  &  13.15  &  0.473  &  1.419  &  0.793   \\
J134739.8+622148.9     &  0.804  &   9.26  &  45.52  &  44.87  &  45.87  &  13.19  &  0.517  &  1.181  &  0.742   \\
J135305.5+044338.3     &  0.523  &   9.51  &  45.26  &  44.95  &  45.95  &  14.94  &  0.514  &  1.293  &  0.778   \\
J142314.2+505537.6     &  0.276  &   8.85  &  44.64  &  44.04  &  45.04  &  14.13  &  0.622  &  1.440  &  0.899   \\
J142606.2+402432.0     &  0.664  &   9.49  &  44.96  &  44.80  &  45.80  &  14.42  &  0.583  &  1.218  &  0.798   \\
J142632.4+025645.6     &  0.861  &   9.37  &  45.62  &  45.33  &  46.33  &  13.88  &  0.538  &  1.097  &  0.727   \\
J150824.7+560423.3     &  0.978  &   9.74  &  45.80  &  45.27  &  46.27  &  14.31  &  0.452  &  1.494  &  0.805   \\
J151526.7+593453.3     &  0.622  &   8.75  &  44.66  &  44.35  &  45.35  &  14.96  &  0.612  &  1.074  &  0.769   \\
J151830.9+483214.5     &  0.576  &   9.53  &  45.54  &  44.91  &  45.91  &  13.69  &  0.470  &  1.430  &  0.796   \\
J152556.2+591659.2     &  0.955  &   9.84  &  46.19  &  45.66  &  46.66  &  13.91  &  0.461  &  1.485  &  0.808   \\
J152942.2+350851.5     &  0.287  &   8.73  &  44.62  &  44.06  &  45.06  &  14.01  &  0.506  &  1.283  &  0.769   \\
J153404.4+580059.4     &  1.572  &   9.88  &  46.73  &  46.44  &  47.44  &  13.83  &  0.274  &  1.538  &  0.702   \\
J154232.0+493842.7     &  0.590  &   8.96  &  44.87  &  44.65  &  45.65  &  14.47  &  0.534  &  1.183  &  0.754   \\
J155853.6+425817.6     &  0.874  &   9.60  &  45.89  &  45.38  &  46.38  &  13.88  &  0.452  &  1.396  &  0.772   \\
J161035.4+480022.3     &  0.247  &   8.26  &  44.33  &  43.42  &  44.42  &  13.68  &  0.413  &  1.313  &  0.718   \\
J161655.6+362134.4     &  2.265  &   9.90  &  47.23  &  46.91  &  47.91  &  14.64  &  0.473  &  1.301  &  0.754   \\
J163856.5+433513.0     &  0.339  &   8.97  &  44.81  &  44.04  &  45.04  &  15.09  &  0.472  &  1.417  &  0.792   \\

\hline
\end{tabular}
\end{center}
\end{table*}
\addtocounter{table}{-1}
%%%%%%%%%%%%%%%%%%%%%%%%%%%%%%%%%%%%%%%%%%%%%%%%%%%%%%%%%%%%%%%%%%%%%%%%%%%%%%%%%%
\begin{table*}
\setlength{\tabcolsep}{0.08in} \caption{Continued\dots
\label{tablesample}}
 \begin{center}
\begin{tabular}{lccccccccccc}
\hline\hline SDSS name &  $z$ & ${\rm Log}~M_{\rm BH}$ & ${\rm
Log}~\nu L_{\nu {\rm peak}}$  &  ${\rm Log}~ L_{\rm BLR}$    &
 ${\rm Log}~ L_{\rm bol}$  &  ${\rm Log}~ \nu_{\rm peak}$    &  $\alpha_{\rm ro}$
  & $\alpha_{\rm ox}$   &  $\alpha_{\rm rx}$   \\
% \hline
    &     & ($M_\odot$)   & ($\rm erg\,s^{-1}$)   & ($\rm erg\,s^{-1}$)  & ($\rm erg\,s^{-1}$)  &  (Hz)        \\
%\hline
(1) & (2) & (3)  & (4)  & (5)  &  (6) & (7) & (8) & (9)  & (10)  \\
\hline
J164054.2+314329.8     &  0.958  &   9.50  &  45.66  &  45.38  &  46.38  &  14.16  &  0.559  &  1.316  &  0.816   \\
J164147.6+393503.4     &  0.539  &   9.04  &  45.18  &  44.68  &  45.68  &  15.95  &  0.499  &  0.986  &  0.664   \\
J164452.6+373009.3     &  0.758  &   9.99  &  46.03  &  45.32  &  46.32  &  13.59  &  0.473  &  1.533  &  0.832   \\
J164544.7+375526.3     &  0.598  &   9.34  &  45.45  &  44.85  &  45.85  &  14.24  &  0.464  &  1.482  &  0.809   \\
J165801.4+344328.5     &  1.939  &   9.85  &  47.41  &  46.69  &  47.69  &  13.71  &  0.523  &  1.624  &  0.896   \\
J170112.4+353353.3     &  0.501  &   8.70  &  44.62  &  44.42  &  45.42  &  14.95  &  0.574  &  0.722  &  0.624   \\
J172051.2+620944.6     &  1.010  &   9.65  &  46.19  &  45.45  &  46.45  &  13.46  &  0.374  &  1.189  &  0.650   \\

\hline
\end{tabular}
\end{center}
\end{table*}
\addtocounter{table}{-2}
%%%%%%%%%%%%%%%%%%%%%%%%%%%%%%%%%%%%%%%%%%%%%%%%%%%%%%%%%%%%%%%%%%%%%%%%%%%%%%%
%%{  Notes: Col. 1: the source name. Col. 2: the redshift. Col. 3: the black hole mass
%%in units of solar mass. Col. 4: the intrinsic synchrotron peak luminosity.  Col. 5:
%%the bolmatric luminosity.  Col. 6: the the intrinsic synchrotron peak frequency,
%%in units of Hz. The Col. 4, 5,  in units of $\rm erg~ s^{-1}$. }

\clearpage

\begin{thebibliography}{99}

\bibitem[Ant\'{o}n \& Browne (2005) ]{anton05}
Ant\'{o}n S., Browne I.~W.~A., 2005, \mnras, 356, 225
\bibitem[Bassani et al.(2007)]{bas07} Bassani L., Landi R., Malizia A., Fiocchi M.T., et al., 2007, \apj, 669, L1
\bibitem[Becker ,White \& Helfand(1995)]{bec95}
Becker R.~H., White R.~L., Helfand D.~J., 1995, \apj, 450, 559
%\bibitem[Blandford \& Znajek(1977)]{BZ1977} Blandford R.D. \& Znajek R.L. 1977 \mnras, 179, 433
%\bibitem[Blandford \& Payne(1982)]{BP1982} Blandford R.D. \& Payne D.G. 1982 \mnras, 199, 883
\bibitem[Boroson \& Green(1992)]{boroson92}
Boroson T. A., Green R.~F., 1992, \apjs, 80, 109
\bibitem[Cao \& Jiang(2001)]{cao01} Cao X., Jiang D. R., 2001, MNRAS, 320, 347
%Cao, X. W. 2003,ApJ, 599, 147
\bibitem[Celotti \& Fabian(1993)]{cel93} Celotti A., Fabian A.~C., 1993, MNRAS, 264, 228
\bibitem[Celotti, Padovani \& Ghisellini(1997)]{cel97}
Celotti A., Padovani P., Ghisellini G., 1997, \mnras, 286, 415
%\bibitem[Celotti, Fabian \& Rees(1998)]{celot98}Celotti A., Fabian A.~C., Rees M.~J., 1998, \mnras, 293, 239
\bibitem[Celotti \& Ghisellini(2008)]{cel08}
Celotti A., Ghisellini G., 2008, \mnras, 385, 283
%\bibitem[Collinge et al.(2005)]{col05} \textbf{Collinge M. J. et al., 2005, \aj, 129, 2542}
\bibitem[D'Elia, Padovani \& Landt(2003)]{del03}
D'Elia V., Padovani P., Landt H., 2003, \mnras, 339, 1081
\bibitem[Dietrich et al.(2002)]{die02}
Dietrich M., Appenzeller I., Vestergaard M., Wagner S. J., 2002, \apj, 564, 581
%\bibitem[Dietrich et al.(2003)]{dietrich03}Dietrich M., Hamann, F., Appenzeller, I.,  Vestergaard, M. 2003,\apj, 596, 817
\bibitem[Dickey \& Lockman(1990)]{dickey90}
Dickey J.~M., Lockman F.~J., 1990, ARA\&A, 28, 215
\bibitem[Forster et al.(2001)]{Forster01}
Forster K., Green P.~J., Aldcroft T.~L., Vestergaard M., Foltz C.~B.,
Hewett P.~C., 2001, \apjs, 134, 35
%\bibitem[Foschini, Ghisellini \& Raiteri(2006)]{fosch06}Foschini L., Ghisellini G., Raiteri C.~M., 2006, \aap, 453,829F
\bibitem[Fossati et al.(1998)]{fossati98}
Fossati G., Maraschi L., Celotti A., Comastri A.,
Ghisellini G., 1998, \mnras, 299, 433
\bibitem[Francis et al.(1991)]{Francis91}
Francis P.~J., Hewett P.~C., Foltz C.~B., Chaffee F.~H.,
Weymann R.~J., Morris S.~L., 1991, ApJ, 373, 465
\bibitem[Frank et al.(2002)]{fra02} Frank J., King A., Raine D., 2002, Accretion Power in Astrophysics
(Cambridge: Cambridge Univ. Press)
\bibitem[Ghisellini et al.(1998)]{ghisel98}
Ghisellini G., Celotti A., Fossati G., Maraschi L., Comastri A.,
1998,  \mnras, 301,451
\bibitem[Ghisellini, Celotti \& Costamante(2002)]{ghi02}
Ghisellini G., Celotti A., Costamante L., 2002, \aap, 386, 833
\bibitem[Ghisellini \& Tavecchio(2008)]{ghisel08}
Ghisellini G., Tavecchio F., 2008, \mnras, 387, 1669
%\bibitem[Giommi et al.(2005)]{giommi05}
%Giommi P., Piranomonte S., Perri M., Padovani P. ,2005, \aap,434, 385
\bibitem[Giommi et al.(2007)]{gio07} Giommi P., Massaro E., Padovani P., Perri M, et al., 2007, \aap, 468, 97
%\bibitem[Giovannini et al.(1988)]{giovan88}
%Giovannini G., Feretti L., Gregorini L., Parma P., 1988, \aap, 199, 73
%\bibitem[Giovannini et al.(2001)]{giova01}
%Giovannini G., Cotton W.~D., Feretti L., Lara L., Venturi T.,
%2001, \apj, 552, 508
\bibitem[Grandi (1982)]{grandi82}
Grandi S.~A., 1982, \apj, 255, 25
\bibitem[Greene \& Ho(2005)]{Green05}
Greene J.~E., Ho L.~C., 2005 \apj, 630, 122
\bibitem[Gregory et al.(1996)]{gre96}
Gregory P.~C., Scott W.~K., Douglas K., Condon J.~J., 1996, \apjs, 103, 427
\bibitem[Gu, Cao \& Jiang (2001)]{gu01} Gu M.~F., Cao X., Jiang D.~R., 2001, \mnras, 327, 1111
\bibitem[Gu et al.(2006)]{gu06} Gu M.~F., Lee C.~U., Pak S., Yim H.~S.,
Fletcher A.~B., 2006, \aap, 450, 39
\bibitem[Gu, Cao \& Jiang(2009)]{gu09}
Gu M.~F., Cao X., Jiang D.~R., 2009, \mnras, arXiv: 0903.1896.
\bibitem[Hu et al.(2008)]{Hu08}
Hu C., Wang J.~M., Ho L.~C., Chen Y.~M., Bian W.~H., Xue S.~J., 2008, \apj, L683, L115
\bibitem[Ivezi\'{c} et al.(2002)]{ive02} Ivezi\'{c} \v{Z}., Menou K., Knapp G. R.,
Strauss M. A., et al., 2002, \aj, 124, 2364
\bibitem[Kaspi et al.(2000)]{kas00} Kaspi S., Smith P.S., Netzer H.,
Maoz D., Jannuzi B.~T., Giveon U., 2000, \apj, 533, 631
\bibitem[Kim et al.(2006)]{kim06}
Kim M., Ho L.~C., Im M., 2006, \apj, 642, 702
\bibitem[Kimball \& Ivezi\'{c}(2008)]{kim08} Kimball A.~E., Ivezi\'{c} \v{Z}., 2008, \aj, 136, 684
\bibitem[Kong et al. (2006)]{kong06}
Kong M.~Z., Wu X.~B., Wang R., Han J.~L., 2006, \chjaa, 6,396
%\bibitem[Landt et al.(2001)]{landt01}
%Landt, H., Padovani, P., Perlman, E. S. 2001, \mnras, 323, 757 %=========        ==???
%\bibitem[Landt et al.(2002)]{lan02} \textbf{Landt H., Padovani P., Giommi P., 2002, \mnras, 336, 945}
%\bibitem[Landt et al.(2004)]{lan04} \textbf{Landt H., Padovani P., Perlman E. S., Giommi P., 2004, \mnras, 351, 83}
\bibitem[Landt et al.(2006)]{lan06} Landt H., Perlman E. S., Padovani P., 2006, \apj, 637, 183
\bibitem[Landt et al.(2008)]{landt08}
Landt H., Padovani P., Giommi P., Perri M., Cheung C.~C., 2008, \apj, 676, 87
\bibitem[Laor(2000)]{lao00} Laor A., 2000, \apj, 543, L111
\bibitem[Lu et al.(2007)]{lu07} Lu Y., Wang T., Zhou H., Wu J., 2007, \aj, 133, 1615
\bibitem[Maiolino et al.(2001)]{mai01} Maiolino R., Salvati M., Marconi A., Antonucci R. R. J., 2001, \aap, 375, 25
\bibitem[Mannucci et al.(2001)]{man01} Mannucci F., Basile F., Poggianti B. M.,
Cimatti A., Daddi E., Pozzetti L., Vanzi L, 2001, \mnras, 326, 745
\bibitem[Maraschi et al.(2008)]{mar08} Maraschi L., Foschini L., Ghisellini G.,
 Tavecchio F., Sambruna R.~M. 2008, \mnras, 391, 1981
%\bibitem[March\~{a} et al.(1996)]{mar96} \textbf{March\~{a} M.J.M., Browne I.W.A., Impey C.D., Smith P.S., 1996, \mnras,
%281, 425}
\bibitem[Martin et al.(2005)]{mar05} Martin D. C., Fanson J, Schiminovich D., Morrissey P., 2005, \apj, 619, L1
\bibitem[McLure et al.(2004)]{mcl04} McLure R. J., Willott C.J., Jarvis M. J., Rawlings S., Hill G. J.,
 Mitchell E., Dunlop J. S., Wold M., 2004, \mnras, 351, 347
%\bibitem[March$\tilde{a}$ et al.(2005)]{browne05}
%March$\tilde{a}$ M.J.M., Browne I.W.A., Jethava N.,
%Ant\'{o}n S., 2005, \mnras, 361, 469
\bibitem[Netzer(1990)]{Netzer90}
Netzer H., 1990 in Active Galactic Nuclei, ed. R. D. Blandford et
al. (Berlin: Springer), 57
\bibitem[Nieppola, Tornikoski \& Valtaoja (2006)]{nieppola06}
Nieppola E., Tornikoski M., Valtaoja E., 2006, \aap, 445, 441
\bibitem[Nieppola et al.(2008)]{nieppola08}
Nieppola E., Valtaoja E., Tornikoski M., Hovatta T., Kotiranta M., 2008, \aap, 488, 867
\bibitem[Padovani (2007)]{padovani07}
Padavani P., 2007, Ap\&SS, 309, 63
\bibitem[Padovani et al.(2002)]{pad02}
Padovani P., Costamante L., Ghisellini G., Giommi P., Perlman E., 2002, \apj, 581, 895
\bibitem[Padovani \& Giommi (1995)]{padovani95}
Padovani P.,  Giommi, P. 1995, \apj, 446, 547
\bibitem[Padovani et al.(1997)]{pad97} Padovani P.,
Giommi P., Fiore F., 1997, Mem. Soc. Astron. Italiana, 68, 147
\bibitem[Padovani et al.(2003)]{padovani03}
Padovani P., Perlman E.~S., Landt H., Giommi P., Perri M., 2003, \apj, 588, 128
\bibitem[Peterson(1997)]{pet97} Peterson B. M., 1997, An Introduction to Active Galactic Nuclei (Cambridge:
Cambridge Univ. Press)
%\bibitem[Padovani \& Perlman et al. (2003)]{padovani03}
%Padovani,P. ; Perlman,E. \&  Landt,H. , \apj, 588, 128
%\bibitem[Perlman et al.(1996)]{per96}
%Perlman E.~S., et al. 1996, \apjs,104, 251
%\bibitem[Perlman et al.(1998)]{perl98}
%Perlman E.~S., Padovani P., Giommi P., Sambruna R., Jones L.~R.,
% Tzioumis A., Reynolds J., 1998, \aj, 115,1253
%\bibitem[Pier et al.(2003)]{pie03} Pier J.~R., Munn J.~A.,
%Hindsley R.~B., Hennessy G.~S., Kent S.~M., Lupton R.~H.,
%Ivezi\'{c} \v{Z}., 2003, \aj, 125, 1559
%\bibitem[Piner et al.(2006)]{pin06}
%Piner B.~G., Bhattarai D., Edwards P.~G., Jones, D.~L., 2006, \apj, 640, 196
%\bibitem[Piner \& Edwards (2004)]{pin04}
%Piner B.~G.,  Edwards P.~G., 2004, \apj, 600, 115
\bibitem[Rawlings \& Saunders(1991)]{raw91}
Rawlings S.~G., Saunders R.~D.~E., 1991, Nature, 349, 138
\bibitem[Rawlings et al.(1989)]{raw89}
Rawlings S.~G., Saunders R.~D.~E., Eales S.~A., Mackay C.~D.,1989, MNRAS, 240, 701
%\bibitem[Rector, Gabuzda \& Stocke(2003)]{rec03}
%Rector T.~A., Gabuzda D.~C., Stocke J.~T., 2003, \aj, 125, 1060
%\bibitem[Rector et al.(2000)]{rec00}
%Rector T.~A., Stocke J.~T., Perlman E.~S., Morris S.~L.,
%Gioia I.~M., 2000, \aj, 120, 1626
%\bibitem[Rector \& Stocke (2001)]{rec01}
%Rector T.~A., Stocke J.~T., 2001, \aj, 122, 565
%\bibitem[Rector \& Stocke (2003)]{re03}
%Rector T.~A., Stocke J.~T., 2003, \aj, 125, 2447
%\bibitem[Sambruna, Maraschi \& Urry(1996)]{sambr96}
%Sambruna R.~M., Maraschi L., Urry M., 1996, \apj, 463, 444
%\bibitem[Savage \& Mathis(1979)]{savage79}
%Savage B.~D., Mathis J.~S. 1979, ARA\&A, 17, 73
\bibitem[Schlegel, Finkbeiner \& Davis(1998)]{sch98}
Schlegel D.~J., Finkbeiner D.~P., Davis M., 1998, \apj, 500, 525
\bibitem[Schneider et al.(2005)]{sch05} Schneider D. P., Hall P. B., Richards G. T.,
 Vanden Berk D. E., et al., 2005, AJ, 130, 367
\bibitem[Shen et al.(2006)]{Shen06}
Shen S.Y., White S.D.M., Mo H.J., Voges W., Kauffmann G.,
Tremonti C., Anderson S.~F., 2006, \mnras, 369, 1639
%\bibitem[Scott et al.(1976)]{scott76}
%Scott R.~L., Leacock R.~J., McGimsey B.~Q., Smith A.~G., Edwards P.~L.,
%Hackney K.~R., Hackney R.~L., 1976, \aj, 81, 7
\bibitem[Skrutskie et al.(2006)]{skr06}
Skrutskie M.~F., Cutri R. M., Stiening R., Weinberg M. D.,  et al.,
2006, \aj, 131, 1163
%\bibitem[Taylor et al.(1994)]{taylor94}
%Taylor G.~B., Vermeulen R.~C., Pearson T.~J., Readhead A.~C.~S.,
%Henstock D.~R., Browne I.~W.~A., Wilkinson P.~N., 1994, \apjs, 95, 345
%\bibitem[Taylor et al. (1996)]{Tay96}
%Taylor G.~B., Vermeulen R.~C., Readhead A.~C.~S., Pearson T.~J.,
%Henstock D.~R., Wilkinson P.~N., 1996,\apjs, 107, 37
\bibitem[Urry \& Padovani (1995)]{urry95}
Urry C.~M., Padovani P., 1995, \pasp, 107 , 803
\bibitem[Vanden Berk et al.(2001)]{vanden01}
Vanden Berk D.~E., Richards G. T., Bauer A.; Strauss M. A., et al.,
2001 \aj ,122, 549
%\bibitem[Veron-Cetty \& Veron (2000)]{veron00}
%Veron-Cetty M.~P., Veron P., 2000, ESOSR, 19, 1
\bibitem[Vestergaard \& Wilkes(2001)]{vestergaard01}
Vestergaard M.,  Wilkes B.~J., 2001, \apjs, 134, 1
\bibitem[Vestergaard \& Peterson(2006)]{Vestergaard06}
Vestergaard M., Peterson B.~M., 2006 \apj, 641, 689
%\bibitem[Voges et al.(1999)]{voges99} Voges W. et al., 1999, \aap, 349, 389
%\bibitem[Voges et al.(2000)]{voges00} Voges W. et al., 2000, IAUC, 7432
\bibitem[Wang, Staubert \& Ho(2002)]{jianmin02}
Wang J.~M., Staubert R., Ho L.~C., 2002, \apj, 579, 554
\bibitem[Wang, Luo \& Ho(2004)]{wan04}
Wang J.~M., Luo B., Ho L.~C., 2004, ApJ, 615, L9
%\bibitem[Wills, Netzer \& Wills(1985)]{wills85}
%Wills B.~J., Netzer H.,  Wills D., 1985, \apj, 288, 94
%\bibitem[Wu et al.(2007)]{wu07}
%Wu, Z.Z.; Jiang, D. R., Gu, M., \& Liu, Y. 2007, \aap, 466, 63
\bibitem[Wu, Gu \& Jiang(2008)]{wu07}
Wu Z.~Z., Gu M.~F., Jiang D.~R., 2009,RAA, 9, 168W
%\bibitem[Xie et al. (2004)]{xie04}
%Xie G.~Z., Zhou S.~B., Li K.~H., Dai H., Chen L.~E., Ma L., 2004, \mnras, 348, 831
%\bibitem[Zirbel \& Baum(1995)]{zirbel}
%Zirbel E.~L., Baum S.~A., 1995, \apj, 448, 521

\end{thebibliography}
\end{document}